\documentclass[aps,prl,twocolumn,groupedaddress,nofootinbib]{revtex4-2}  
\usepackage{graphicx}  
\usepackage{dcolumn}   
\usepackage{bm}        
\usepackage{amssymb,amsfonts,amsmath,physics}   
\usepackage{slashed}
\usepackage{subfig}
\usepackage{graphicx}
\usepackage{lipsum}
\usepackage{multirow}
\usepackage{chngcntr}
\usepackage{slashed}
\PassOptionsToPackage{demo}{graphicx}
\usepackage{framed}

\usepackage{tikz-feynman}
\tikzfeynmanset{compat=1.1.0}
\usetikzlibrary{arrows}
\tikzset{ vertex/.style={circle,draw, minimum size=1.5em}, edge/.style={->,> = latex'}}

\usepackage[bookmarks, breaklinks, colorlinks, urlcolor=blue, citecolor=red, linkcolor=blue]{hyperref}

\usepackage[normalem]{ulem}

\newif\ifstartedinmathmode
\newcommand\encircled[1]{%
  \relax\ifmmode\startedinmathmodetrue\else\startedinmathmodefalse\fi%
  \tikz[baseline,anchor=base]{%
  \node[draw,circle,outer sep=0pt,inner sep=.2ex]
    {\ifstartedinmathmode$#1$\else#1\fi};}%
}
\newcommand*\circi[1]{\tikz[baseline=(char.base)]{\node[shape=circle,fill=none,draw=black,path fading=none,inner sep=1.7pt] (char) {\footnotesize #1};}}
\newcommand*\circii[1]{\tikz[baseline=(char.base)]{\node[shape=circle,fill=none,draw=black,path fading=none,inner sep=1.0pt] (char) {\footnotesize #1};}}
\newcommand*\circiii[1]{\tikz[baseline=(char.base)]{\node[shape=circle,fill=none,draw=black,path fading=none,inner sep=0pt] (char) {\footnotesize #1};}}
\newcommand*\circiv[1]{\tikz[baseline=(char.base)]{\node[shape=circle,fill=none,draw=black,path fading=none,inner sep=0.3pt] (char) {\footnotesize #1};}}

\newcommand\figstrut[2]{
  \dimen0=#1%
  \advance\dimen0 by -#2%
  \divide\dimen0 by -2%
  \dimen1=#1%
  \advance\dimen1 by \dimen0%
  \vrule height \dimen1 depth \dimen0 width 0pt\relax%
}

\def\to {\rightarrow}

\definecolor{deepmagenta}{rgb}{0.8, 0.0, 0.8}
\definecolor{mediumtealblue}{rgb}{0.0, 0.33, 0.71}
\definecolor{warmblack}{rgb}{0.0, 0.26, 0.26}
\definecolor{bostonuniversityred}{rgb}{0.8, 0.0, 0.0}
\definecolor{junglegreen}{rgb}{0.16, 0.67, 0.53}
\definecolor{lightcornflowerblue}{rgb}{0.6, 0.81, 0.93}
\newcommand{\bmt}{\begin{pmatrix}}
\newcommand{\emt}{\end{pmatrix}}
\newcommand{\ba}{\begin{array}{c}}
\newcommand{\ea}{\end{array}}
\newcommand{\be}{\begin{equation}}
\newcommand{\ee}{\end{equation}}
\newcommand{\bea}{\begin{eqnarray}}
\newcommand{\eea}{\end{eqnarray}}

\newcommand{\bi}{\begin{itemize}}
\newcommand{\ei}{\end{itemize}}

\newcommand{\baz}{\begin{array}{cc}}
\newcommand{\mathsym}[1]{{}}

\newcommand{\bt}{\begin{tabular}}
\newcommand{\et}{\end{tabular}}

\newcommand{\benu}{\begin{enumerate}}
\newcommand{\eenu}{\end{enumerate}}

\newcounter{mysfig}
\counterwithin{mysfig}{figure}

\renewcommand\themysfig{1\alph{mysfig}}
\makeatletter
\newcommand\Scaption[1]{%
\refstepcounter{mysfig}%
\vskip.01\abovecaptionskip
  \sbox\@tempboxa{\small\themysfig~#1}%
  \ifdim \wd\@tempboxa >\hsize
    \small\themysfig~#1\par
  \else
    \global \@minipagefalse
    \hb@xt@\hsize{\hfil\box\@tempboxa\hfil}%
  \fi
  \vskip\belowcaptionskip}
\makeatother

\begin{document}


\title{Dynamics of the pseudo-{\tt FIMP} in presence of a thermal dark matter}

\author{Subhaditya Bhattacharya}
\email[E-mail: ]{subhab@iitg.ac.in}
\affiliation{Department of Physics, Indian Institute of Technology Guwahati, Assam-781039, India}

\author{Jayita Lahiri}
\email[E-mail: ]{jayita.lahiri@desy.de}
\affiliation{Institut f{\"u}r Theoretische Physik, Universit{\"a}t Hamburg, 22761 Hamburg, Germany} 

\author{Dipankar Pradhan} 
\email[E-mail: ]{d.pradhan@iitg.ac.in}
\affiliation{Department of Physics, Indian Institute of Technology Guwahati, Assam-781039, India}


\begin{abstract} 
In a two-component dark matter (DM) set up, when DM$_1$ is in equilibrium with the thermal bath, the other DM$_2$ can 
be equilibrated just by sizeable interaction with DM$_1$, even without any connection with the visible-sector particles. 
We show that such DM candidates (DM$_2$) have unique `freeze-out' characteristics impacting the relic density, direct and collider 
search implications, and propose to classify them as `pseudo-{\tt FIMP}' (p{\tt FIMP}). The dynamics of p{\tt FIMP} is studied in a model-independent manner
by solving generic coupled Boltzmann Equations (cBEQ), followed by a concrete model illustration. 
\end{abstract}

\maketitle

\noindent
Albeit tantalizing astrophysical and cosmological evidences, what constitutes the dark matter (DM) component 
of the universe remains a mystery with no direct experimental evidence in terrestrial experiments so far. 
Amongst DM geneses, thermal {\em freeze-out} of Weakly Interacting Massive Particle ({\tt WIMP}) \cite{Kamionkowski:1990ni,Jungman:1995df} 
and non-thermal {\em freeze-in} of Feebly Interacting Massive Particle ({\tt FIMP}) \cite{Hall:2009bx} can both address the observed DM relic density 
$\rm\Omega_{\rm DM} h^2 \sim 0.12\pm .0012$ \cite{Planck:2018vyg} with widely different interaction strengths with the Standard Model (SM) particles. 
Multipartite dark sectors offer interesting cosmological and phenomenological consequences, like modified freeze-out \cite{Bhattacharya:2013hva}, 
relieving direct search constraints \cite{Bhattacharya:2016ysw}, addressing small scale structure anomalies of the universe~\cite{Ghosh:2021wrk} etc. 
The distinguishability of a two-component {\tt WIMP} scenario, from that of single-component one, in terms of a double-hump in the missing energy distribution at collider~\cite{Bhattacharya:2022wtr,Bhattacharya:2022qck} or a kink in the recoil energy spectrum~\cite{Herrero-Garcia:2017vrl,Herrero-Garcia:2018qnz} of the direct search experiment have been studied.

When the interaction between a {\tt WIMP} and a {\tt FIMP} remains `feeble', the {\tt FIMP} continues to be out of equilibrium 
\cite{Belanger:2021lwd,Bhattacharya:2021rwh,Costa:2022oaa}, but enhanced interaction (of weak strength) can make the {\tt FIMP} 
equilibrate to thermal bath and undergo {\em freeze-out} with some unique features, which are essentially model-independent, but relies 
on the properties of its thermal DM partner. We propose to classify such a particle into a category called `pseudo-{\tt FIMP}' (p{\tt FIMP}), 
as it is a thermal DM in spite of having feeble connection with the SM. Also, p{\tt FIMP} can be realized in presence of {\it any} thermal DM like 
Strongly Interacting Massive Particle ({\tt SIMP}), independent of the depletion mechanism. 
Some model-dependent studies which mimic a p{\tt FIMP}-like situation, have been studied \cite{Belanger:2011ww,Bhattacharya:2013hva,Maity:2019hre,DiazSaez:2021pfw}, 
but without elaborating the generic features that such particles possess. This letter illustrates the 
p{\tt FIMP} characteristics in a model-independent way followed by a concrete model illustration and detection possibilities. 
Actually the idea of p{\tt FIMP} has been conceptualised and demonstrated here for the first time, that occurs in a particular limit of 
the DM-DM and DM-SM interactions in a multicomponent set up. This leads to several model possibilities to be explored in the 
p{\tt FIMP} limit as detailed in \cite{Bhattacharya:2022vxm}.

The evolution of DM number density in {\tt WIMP}-{\tt FIMP} framework is governed by a set of coupled Boltzmann equations (cBEQ) as shown in Eq.~\eqref{eq:cbeq}, 
in terms of yield ($Y=\frac{n}{s}$) as a function of $x=\mu_{12}/T$, where $\mu_{12}=m_1 m_2/\left(m_1+m_2\right)$ denotes the 
reduced mass of the two DMs, and $T$ denotes the temperature of the thermal bath.

\begin{eqnarray}\begin{split}
 \frac{dY_1}{dx}=&-\frac{{s}}{x~ H(x)}\Biggl[\langle\sigma v\rangle_{11\to\rm{SM}}\Bigl(Y_1^2-Y_{1}^{\rm{eq}^2}\Bigr)\\&+\langle\sigma v\rangle_{11\to22}\Bigl(Y_1^2-\frac{Y_1^{\rm eq^2}}{Y_2^{\rm eq^2}}Y_2^2\Bigr)\Biggr],\\
\frac{dY_2}{dx}=&\frac{2~{s}}{x~ H(x)}\Biggl[\dfrac{1}{{s}}\langle\Gamma_{\rm SM\to22}\rangle (Y_{\rm SM}^{\rm eq }-\frac{Y_2^2}{Y_2^{\rm eq^2}}Y_{\rm SM}^{\rm eq})\\&+\langle\sigma v\rangle_{\rm{SM}\to 22}\Bigl(Y_{\rm SM}^{\rm eq^2}-\frac{Y_2^2}{Y_2^{\rm eq^2}}Y_{\rm SM}^{\rm eq^2}\Bigr)\\&+\langle\sigma v\rangle_{11\to22}\Bigl(Y_1^2-\frac{Y_{1}^{\rm eq^2}}{Y_{2}^{\rm eq^2}}Y_2^2\Bigr)\Biggr]\,.\end{split}\label{eq:cbeq}
\end{eqnarray}

In the above equation, ${s}\,=\,\frac{ 2\pi^2}{45}g_{*}^{{s}}\left(\frac{\mu_{12}}{x}\right)^3~,~H(x)\,=\,1.67\sqrt{g_*^{\rho}}{\mu^2_{12}}x^{-2}\rm M_{pl}^{-1}$.
Subscripts $1,~2$ in Eq.~\eqref{eq:cbeq} and in the rest of the draft denote {\tt WIMP} and {\tt FIMP} (p{\tt FIMP}) components respectively. 
Interactions that crucially determine the DM densities are {\tt WIMP} depletion to SM governed by 
$\langle\sigma v\rangle_{11\to\rm{SM}}$, {\tt FIMP} production from the SM bath via annihilation 
$\langle\sigma v\rangle_{\rm{SM}\to 22}$ and/or decay $\langle\Gamma_{\rm SM\to22}\rangle$, 
and {\tt WIMP}-{\tt FIMP} conversion $\langle\sigma v\rangle_{11\to22}$ ($\langle .. \rangle$ denotes thermal average and $v$ is M$\ddot{o}$llar velocity). 
The cBEQ for a two component {\tt WIMP} case is also the same as Eq.~\ref{eq:cbeq}. 
The difference lies in the $(i)$ strength of the DM-SM interactions; $\langle\sigma v\rangle_{\tt {\tt WIMP}} \sim 10^{-8}~{\rm GeV}^{-2}$, whereas 
$\langle\sigma v\rangle_{\tt {\tt FIMP}} \sim 10^{-20}~{\rm GeV}^{-2}$ and $(ii)$
initial conditions, $Y_{\tt {\tt WIMP}}|_{x \sim 0}=Y^{\rm eq}\sim x^{3/2}e^{-x}$, $Y_{\tt {\tt FIMP}}|_{x \sim 0}=0$. 
In the following, the conversion $\langle\sigma v\rangle_{11\to22}$ is varied from `feeble' to `weak' strength to show the transition from 
{\tt FIMP} to p{\tt FIMP} state. Other notations in Eq. \eqref{eq:cbeq} are all standard and available in any DM text \cite{Bhattacharya:2016ysw}.


For solving Eq.~\eqref{eq:cbeq} to obtain DM yields in a model-independent way, we choose some benchmark 
values of the DM masses and $\langle\sigma v\rangle$. As the temperature dependence in $\langle\sigma v\rangle$ 
requires $\sigma(s)$ ($s$ is c.o.m energy) i.e. diagrams that contribute to annihilation, in absence of which we assume here $\langle\sigma v\rangle \approx \sigma^{\rm T}$; 
a temperature-independent threshold value representing $(\sigma v)$ at the {\em freeze-out (freeze-in)} temperature for {\tt WIMP}/p{\tt FIMP} ({\tt FIMP}). 
Full $\langle\sigma v\rangle$ is used for a model-specific analysis and the dynamical features remain the same.

\begin{figure*}[hptb!]
\begin{minipage}[t]{\columnwidth}
\begin{minipage}[b]{\columnwidth}
\includegraphics[width=\linewidth]{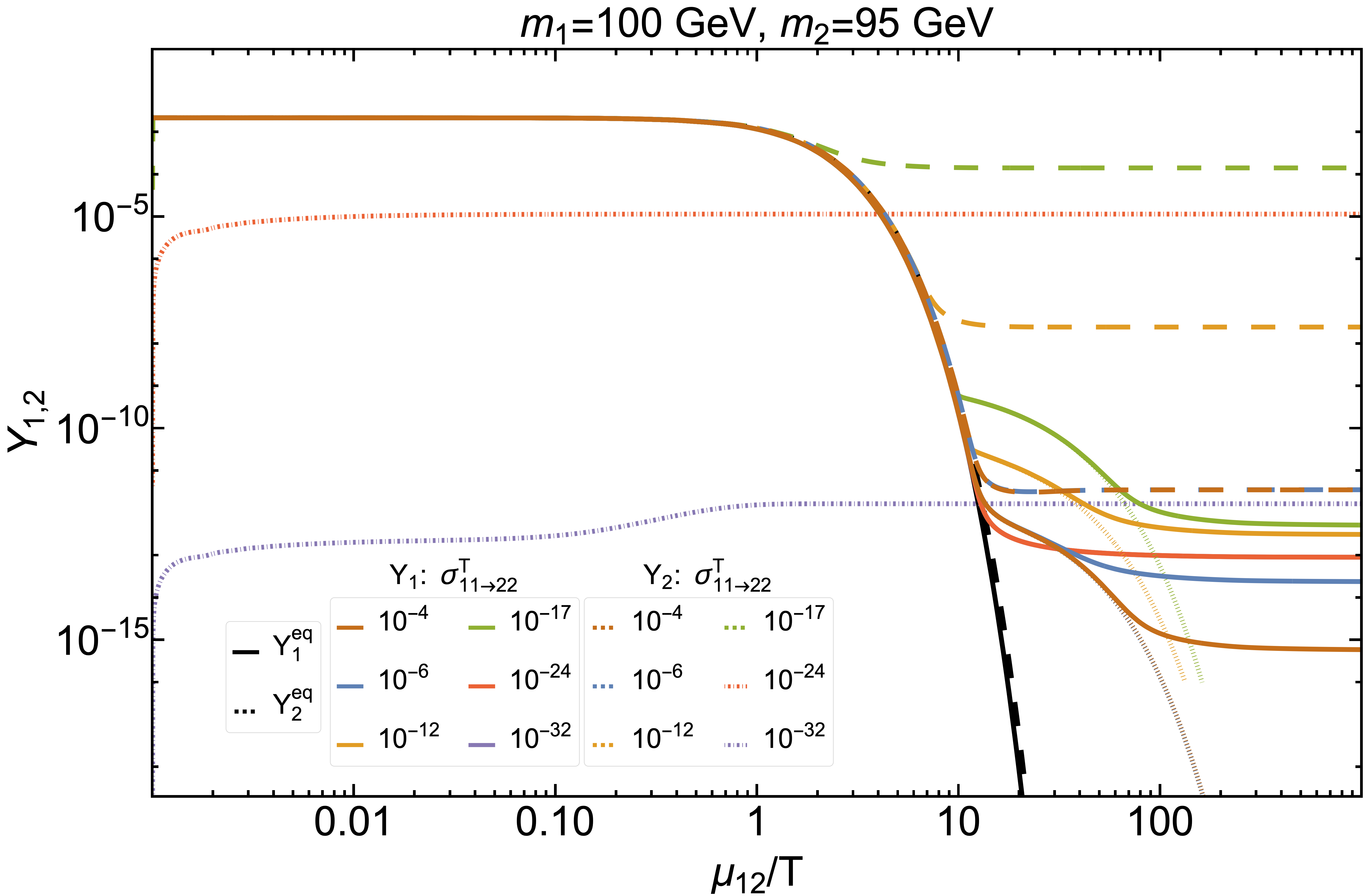}\par
\end{minipage}
\Scaption{}
\label{fig:a}
\end{minipage}\hfill
\begin{minipage}[t]{\columnwidth}
\begin{minipage}[b]{\columnwidth}
\includegraphics[width=\linewidth]{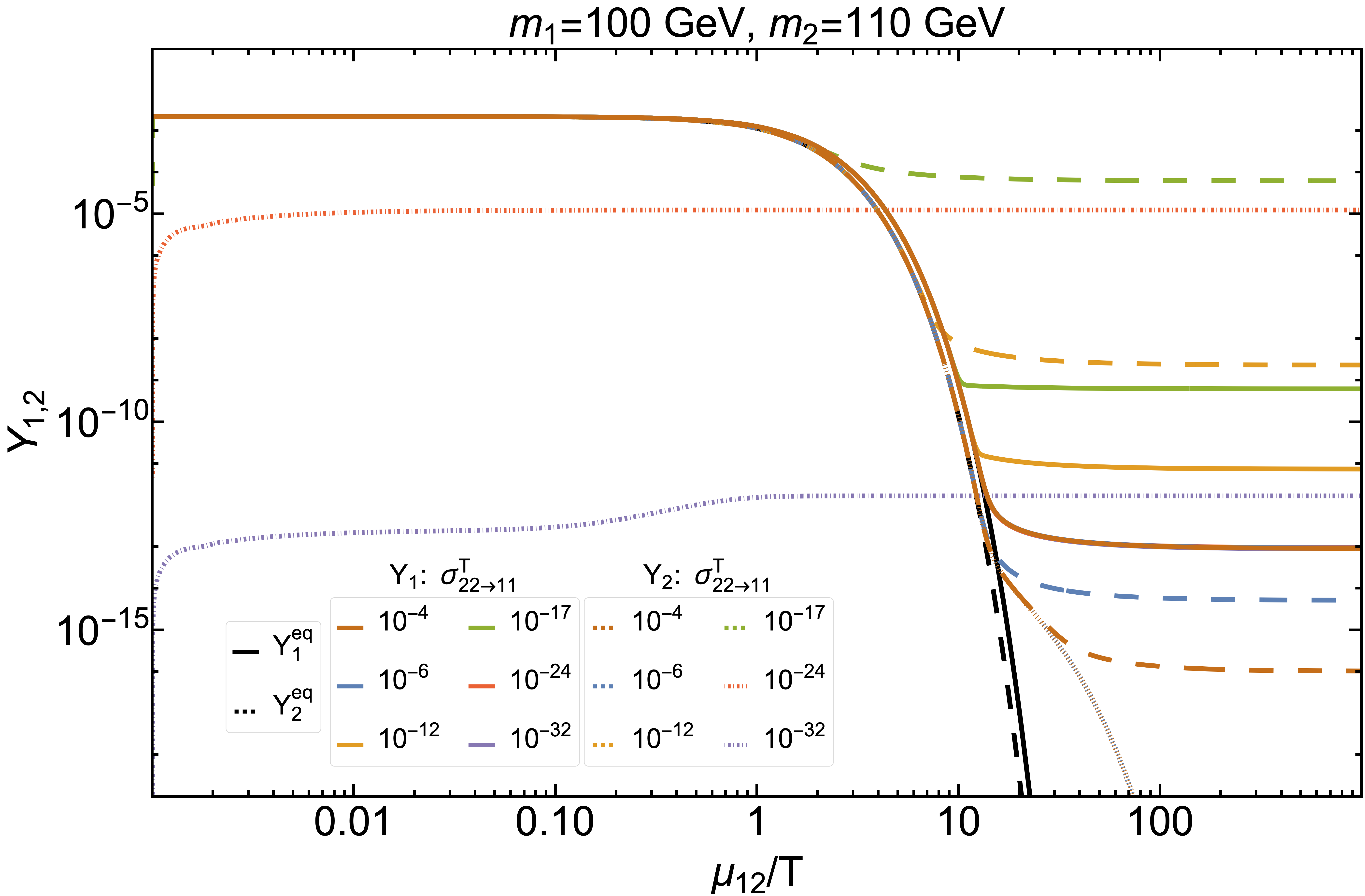}
\end{minipage}
\Scaption{}
\label{fig:b}
\end{minipage}%
\vspace{0.5cm}
\begin{minipage}[t]{\columnwidth}
\begin{minipage}[b]{\columnwidth}
\includegraphics[width=\linewidth]{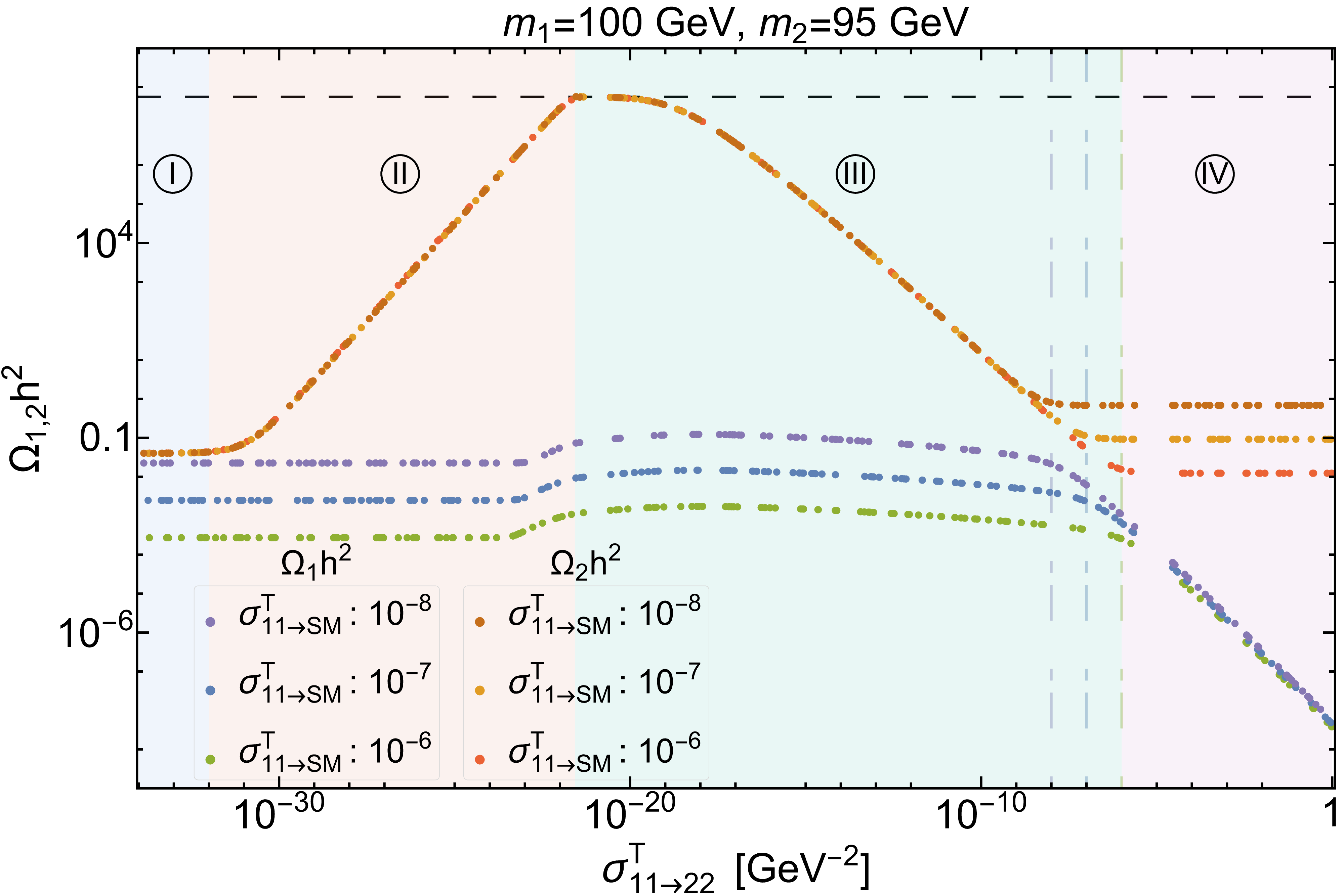}\par
\end{minipage}
\Scaption{}
\label{fig:c}
\end{minipage}\hfill
\begin{minipage}[t]{\columnwidth}
\begin{minipage}[b]{\columnwidth}
\includegraphics[width=\linewidth]{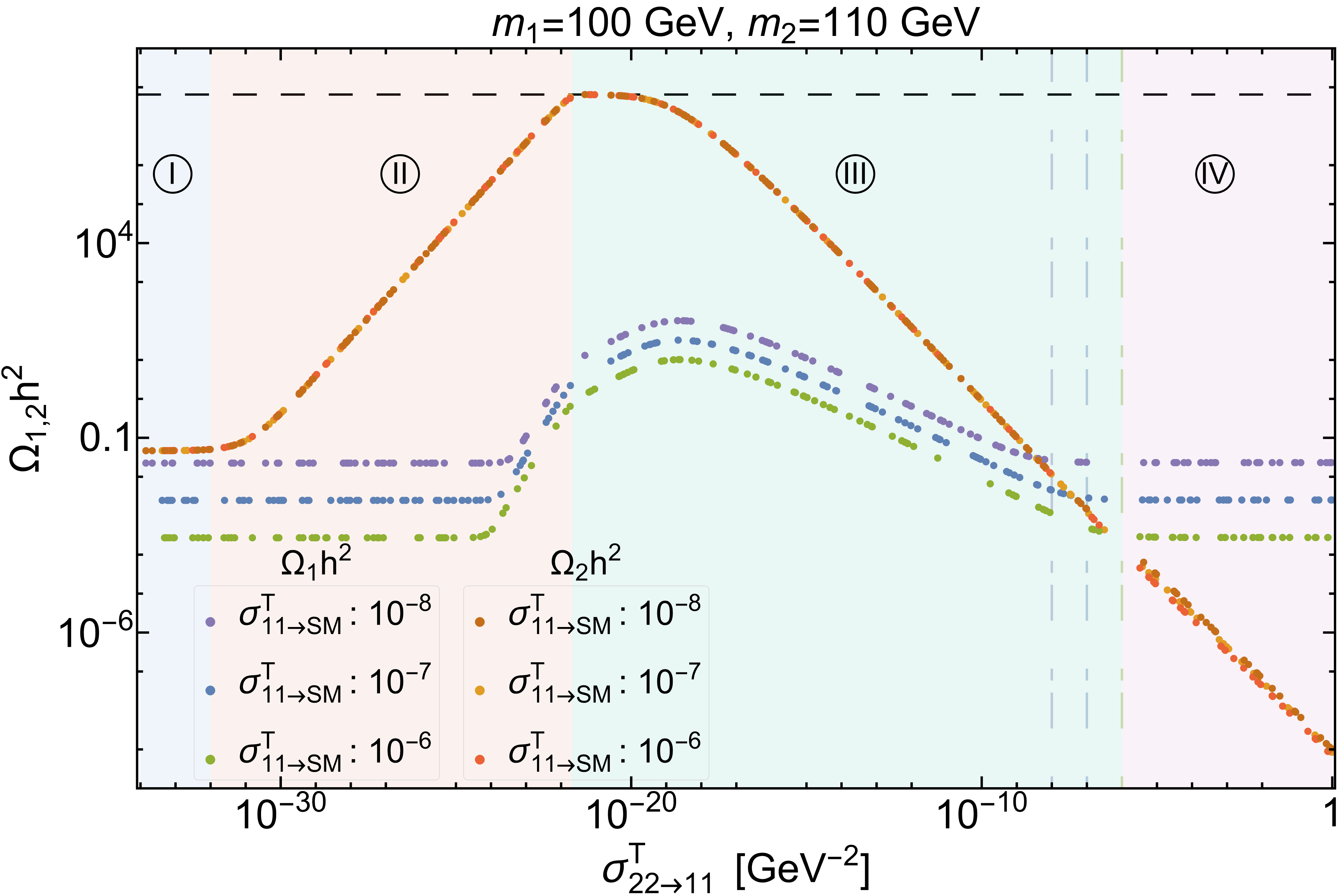}
\end{minipage}
\Scaption{}
\label{fig:d}
\end{minipage}
\caption{(a) $Y_{1,2}(x)$ for mass hierarchy 1, where solid (dotted) [dashed] curves denote {\tt WIMP} ({\tt FIMP}) [p{\tt FIMP}] with 
equilibrium distributions marked in black; (b) Same as (a) for hierarchy 2; (c) ${\rm\Omega_{1,2} h^2}$ as a function of $\sigma^{\rm T}_{11\rightarrow 22}$ for hierarchy 1; 
red/yellow/orange dotted lines represent {\tt FIMP}(p{\tt FIMP}), while green/blue/pink ones represent {\tt WIMP}; (d) same as (c) for hierarchy 2. 
For all the plots $\rm~\Gamma^T_{SM\to22}=10^{-23}GeV^{-1}$, $\rm~\sigma^{\rm T}_{SM\to 22}=10^{-32}GeV^{-2}$. 
For upper panel plots, we choose $\rm \sigma^T_{11\to SM}= 10^{-7}GeV^{-2}$. See figure inset/caption for other parameters kept fixed.}
\label{fig:mod-ind-yld}
\end{figure*}

Plots in fig.~\ref{fig:mod-ind-yld} summarise the main outcome of the model independent analysis. In figures \ref{fig:a}, \ref{fig:b}, 
$Y_{1,2}$ are plotted as a function of $x$, for different conversion rates. In figures \ref{fig:c}, \ref{fig:d}, ${\rm{ \Omega_{1,2}h^2}}$ 
are plotted as a function of $\sigma^{\rm T}_{11 \to 22}$ (left), $\sigma^{\rm T}_{22 \to 11}$ (right). 
Note, that $\langle \sigma v\rangle_{11 \to 22}=\langle \sigma v\rangle_{22 \to 11} \rm(Y_2^{eq}/Y_1^{eq})^2$. 
All the cross-sections are in the units of GeV$^{-2}$ and in the following text we may omit writing it explicitly. 
Both the mass hierarchies (1) $m_1>m_2$ (on left) and (2) $m_1<m_2$ (on right) are shown. In hierarchy (1), conversion from {\tt WIMP} to {\tt pFIMP}
is kinematically allowed, while the reverse process is Boltzmann suppressed ($\sim e^{-2\delta m~x}, ~\delta m=m_1-m_2$). 
For hierarchy (2), its the other way round. The whole analysis can be divided into four different regions of varying conversion cross-section, 
\circi{I} : $\sigma^{\rm T}_{11 \to 22}<\sigma^{\rm T}_{\rm{SM}\to 22}$, \circii{II} : 
$\sigma^{\rm T}_{\rm{SM}\to 22}\lesssim\sigma^{\rm T}_{11 \to 22}\ll \sigma^{\rm T}_{11\to\rm{SM}} $, \circiii{III} : 
$\sigma^{\rm T}_{\rm{SM}\to 22}\ll\sigma^{\rm T}_{11 \to 22}< \sigma^{\rm T}_{11\to\rm{SM}}$, 
\circiv{IV} : $ \sigma^{\rm T}_{11\to\rm{SM}}\lesssim\sigma^{\rm T}_{11 \to 22}$, as shown by the colour bars in figs. \ref{fig:c}, \ref{fig:d}.

In region \circi{I}, when the conversion rate is very small ($\sigma^{\rm T}_{11 \to 22} \sim 10^{-32}$) it has no role in {\tt FIMP} or {\tt WIMP} relic density. 
The {\tt FIMP} remains out-of-equilibrium and freeze-in occurs via $\sigma^{\rm T}_{\rm{SM}\to 22}, \Gamma^{\rm T}_{\rm SM\to22}$, see 
the blue dotted lines in figs. \ref{fig:a}, \ref{fig:b} and horizontal red/yellow/orange dotted lines in figs. \ref{fig:c}, \ref{fig:d}. 

In region \circii{II}, when $\sigma^{\rm T}_{11 \to 22} \gtrsim \sigma^{\rm T}_{\rm{SM}\to 22}$, the {\tt FIMP} is still out of 
equilibrium but the freeze-in is guided by the conversion process, see the red dotted lines in figs. \ref{fig:a}, \ref{fig:b} 
($\sigma^{\rm T}_{11 \to 22} \sim 10^{-24}$) and the growing red/yellow/orange dotted lines with larger $\sigma^{\rm T}_{11 \to 22}$ in figs. \ref{fig:c}, \ref{fig:d}. 
The {\tt WIMP} still remains unaffected, evident from the red/violet solid lines in figs. \ref{fig:a}, \ref{fig:b}, 
and horizontal blue/green/violet dotted lines in figs. \ref{fig:c}, \ref{fig:d}.

In region \circiii{III}, as the conversion rate $\sigma^{\rm T}_{11\to 22}$ is increased further, 
the {\tt FIMP} starts following the thermal distribution, see green ($\sigma^{\rm T}_{11\to 22}=10^{-17}$) and
orange ($\sigma^{\rm T}_{11\to 22}=10^{-12}$) dashed lines in figs. \ref{fig:a}, \ref{fig:b}. This is when the {\tt FIMP} turns into p{\tt FIMP} and 
undergoes ``freeze-out''.  Larger conversion cross-section keeps p{\tt FIMP} longer in the thermal bath resulting smaller relic density; 
see the steady decline in red/yellow/orange dotted lines with larger $\sigma^{\rm T}_{11\to 22}(\approx 10^{-17}$) in figs. \ref{fig:c}, \ref{fig:d}. 
Notably, the freeze-out point of p{\tt FIMP} is governed purely by its conversion rate from (to) the {\tt WIMP}. 
However, the {\tt WIMP} freeze-out is additionally dictated by $\sigma^{\rm T}_{11\to \rm SM}$. Therefore p{\tt FIMP} always decouples 
before or together with the {\tt WIMP} (compare dashed and solid lines in figs. \ref{fig:a}, \ref{fig:b}) and provides an important difference between 
{\tt WIMP}-{\tt WIMP} and p{\tt FIMP}-{\tt WIMP} case. For heavier {\tt WIMP} (hierarchy 1), the {\tt WIMP} freeze-out
shows a bump (see fig. \ref{fig:a}), as it attains a modified equilibrium distribution following Eq.~\eqref{modify-eq}, after p{\tt FIMP} decouples 
\cite{Bhattacharya:2013hva}, 

\begin{align}\rm n_{1}^{\rm eq^{\prime}}= n_{1}^{\rm eq}\left(\frac{\sigma^T_{11\to SM}+\sigma^T_{11\to 22}\left(n_{2}/n_{2}^{eq}\right)^2}{\sigma^T_{11\to SM}+\sigma^T_{11\to 22}}\right)^{1/2}\, .\label{modify-eq}\end{align}

When p{\tt FIMP} is heavier (hierarchy 2), the conversion from {\tt WIMP} to p{\tt FIMP} is kinematically suppressed, and no modified 
freeze-out pattern for {\tt WIMP} is observed (see fig.~\ref{fig:b}). This provides another distinction of p{\tt FIMP}-{\tt WIMP} scenario from the {\tt WIMP}-{\tt WIMP} one.   
After {\tt FIMP} equilibrates to thermal bath, larger production of {\tt WIMP} from p{\tt FIMP} enhances 
{\tt WIMP} relic density, see the jump from red thick line ($\sigma^{\rm T}_{11\to 22}=10^{-24}$) to green thick line 
($\sigma^{\rm T}_{11\to 22}=10^{-17}$) in figs. \ref{fig:a}, \ref{fig:b}. This feature is also observed in the
violet/blue/green dashed lines in figs \ref{fig:c}, \ref{fig:d}. For hierarchy (1), the departure of the {\tt WIMP} 
from the original equilibrium distribution to modified equilibrium (Eq.~\ref{modify-eq}) depends on $\sigma^{\rm T}_{11\to 22}$ 
(compare green and orange thick lines in fig. \ref{fig:a}), however the freeze-out from modified equilibrium is primarily governed by 
$\sigma^{\rm T}_{11\to \rm SM}$, yielding same $\Omega_{1}{\rm h}^2$ for different $\sigma^{\rm T}_{11\to 22}$. 
On the contrary, for hierarchy (2), when p{\tt FIMP} is heavier, the {\tt WIMP} production from p{\tt FIMP} 
is substantial due to kinematic accessibility, as a result the {\tt WIMP} freeze out is delayed, and the {\tt WIMP} yield goes down with 
enhanced conversion rate (see fig. \ref{fig:d}), providing a crucial hierarchical distinction.

In region \circiv{IV}, with $\sigma^{\rm T}_{11 \to 22}> \sigma^{\rm T}_{11\to\rm{SM}}$, we see even more interesting consequences. 
For hierarchy (1), the {\tt WIMP} relic density falls drastically with larger $\sigma^{\rm T}_{11 \to 22}$ due to larger depletion (violet/green/blue dotted 
lines in fig.~\ref{fig:c}), while the p{\tt FIMP} yield remains the same, and is dependent only on $\sigma^{\rm T}_{11\to \rm SM}$ 
(horizontal brown/yellow/red dotted lines in fig.~\ref{fig:c}, or overlapping dashed lines in fig.~\ref{fig:a} for $\sigma^{\rm T}_{11\to 22}=10^{-6}, 10^{-4}$). 
This is because p{\tt FIMP} yield in hierarchy (1) is governed by two quantities, {\tt WIMP} 
number density and conversion rate. Now, the {\tt WIMP} number density falls with larger $\sigma^{\rm T}_{11\to 22}$, while that is compensated 
by the enhanced production via $\sigma^{\rm T}_{11\to 22}$ which keeps the p{\tt FIMP} yield almost constant in this region.  
For hierarchy (2), the situation is however different. With increasing conversion, p{\tt FIMP} yield decreases significantly, 
while the {\tt WIMP} yield remains almost constant (see fig.~\ref{fig:d}). This is attributed again to the combined effect of 
decreasing p{\tt FIMP} number density along with larger {\tt WIMP} production from p{\tt FIMP} with larger conversion rate. 
Since p{\tt FIMP} has no SM interaction, it goes out of the original equilibrium distribution as
{\tt WIMP} freezes out. However, due to considerable conversion to {\tt WIMP}, it achieves a modified equilibrium distribution as seen from the 
orange and blue dashed lines in fig. \ref{fig:b}. 

\begin{table*}{\footnotesize
\begin{ruledtabular}
\begin{tabular}{ccccccccc}
$\rm DM_1-DM_2$ & $\langle\sigma v_{\rm SM-DM_1}\rangle$ & $ \langle\sigma v_{\rm SM-DM_2}\rangle$  & $ \langle\sigma v_{\rm DM_1-DM_2}\rangle$ & Observed Relic & Direct-Detection&Indirect-Detection \\
\hline
{\tt WIMP}-{\tt WIMP} & Weak & Weak & Weak/Feeble & Yes~\cite{Bhattacharya:2016ysw} &Yes-Yes~\cite{Bhattacharya:2018cgx,Bhattacharya:2016ysw}& Yes-Yes\cite{Yaguna:2021rds} \\
~~~~~{\tt WIMP}-p{\tt FIMP}~~~~ & Weak & Feeble & Weak & ~~~Yes~\cite{Bhattacharya:2013hva,DiazSaez:2021pfw} & ~~~Yes-Yes\cite{Bhattacharya:2013hva}~~~~~~~~& Yes-Yes~~~~~\\ 
~~~~~{\tt FIMP}-{\tt FIMP}~~~~ & Feeble & Feeble & Weak/Feeble &~~~~ Yes~\cite{Pandey:2017quk,Abdallah:2020fpb,Ghosh:2021wrk}& No-No~\cite{Ghosh:2021wrk}~~~~~&~No-No~~~~~\
\end{tabular}
\end{ruledtabular}}
\caption{Generic two-component DM scenarios, order of the interaction with SM as well as between DMs, possibilities of satisfying relic abundance and being probed in direct (indirect) search experiments.}
\label{tab:possible-scenario}
\end{table*}

\begin{figure}[hptb!]
\includegraphics[width=1\linewidth]{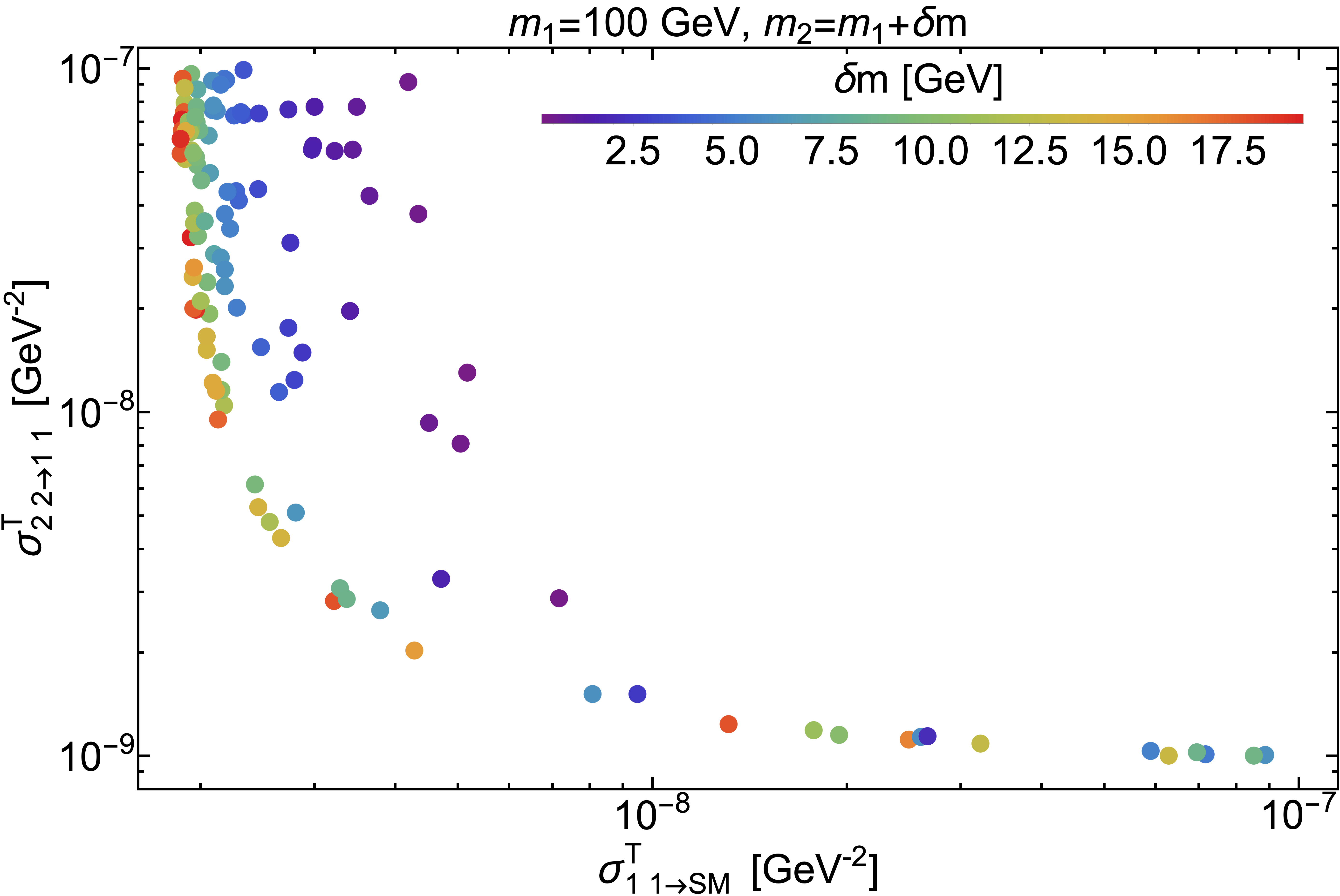}
\caption{Relic density allowed points in the $\sigma^{\rm T}_{11\to {\rm SM}}-\sigma^{\rm T}_{22\to 11}$ plane for $m_2 > m_1$ in {\tt WIMP}-p{\tt FIMP} limit. $\delta m=|m_1 - m_2|$ is varied as shown in the colour-bar. We choose $\Gamma_{{\rm SM}\to 22}^{{\rm T}} =10^{-20}$ GeV$^{-1},~ \sigma^{\rm T}_{{\rm SM} \to 22} =10^{-30}$ GeV$^{-2}$, $m_1=100$ GeV and decaying particle mass $\rm m_s = 300$ GeV.}
\label{fig:relic_scan-mi}
\end{figure}

We now perform a model-independent scan of the {\tt WIMP}-p{\tt FIMP} parameter space, where we assume that annihilation 
cross-sections and masses have adequate freedom to be varied independently.  
In fig.~\ref{fig:relic_scan-mi}, we scan points that satisy relic density 
($\rm \Omega_{ 1} h^2+\Omega_{ 2} h^2 = 0.12\pm .0012$) in $\sigma^{\rm T}_{11\to \rm{SM}}-\sigma^{\rm T}_{22\to 11}$ 
plane for hierarchy (2). {\tt WIMP} mass is kept fixed at $100~\rm GeV$ and the mass splitting between {\tt WIMP} and p{\tt FIMP} $\delta m=|m_1 - m_2|$ 
is varied as shown in the colour bar. When the conversion rate is small, $\sigma^{\rm T}_{11\to \rm{SM}}$ requires to be large in order 
to achieve {\tt WIMP} density within the limit. Understandably $\delta m$ does not play a major role here, and a large range of $\delta m$ is 
allowed (right side of fig. \ref{fig:relic_scan-mi}). With larger conversion rate, smaller $\sigma^{\rm T}_{11\to \rm{SM}}$ is required 
(left of fig. \ref{fig:relic_scan-mi}). If we decrease $\sigma^{\rm T}_{11\to \rm{SM}}$ keeping $\sigma^{\rm T}_{22\to 11}$ fixed, 
the {\tt WIMP} tends to become overabundant, unless the conversion contribution from p{\tt FIMP} to {\tt WIMP} is decreased by increasing 
$\delta m$ via Boltzmann suppression. However, for hierarchy (1), the situation is different and the whole 
$\sigma^{\rm T}_{11\to \rm{SM}}-\sigma^{\rm T}_{11\to 22}$ parameter space within $\delta m=|m_1 - m_2|\lesssim 10$ GeV becomes allowed. 

In summary, the allowed points accommodate mostly small mass difference 
($\delta m \lesssim 10$ GeV) and the mass hierarchy between {\tt WIMP} and p{\tt FIMP} play a crucial 
role in the consequent DM phenomenology, absent in usual {\tt WIMP}-{\tt WIMP} set up. Here we have ensured that the conversion cross-section 
remains well below the bullet cluster observational data~\cite{Clowe:2006eq} that suggests the upper limit on the 
self-interaction cross-section of DM(s) per unit mass to be, $\rm{\sigma/m<0.7~cm^2 g^{-1}\sim 3.2\times 10^3~GeV^{-3}}$.

\begin{figure}[htb!] 
\centering
\includegraphics[width=1\linewidth]{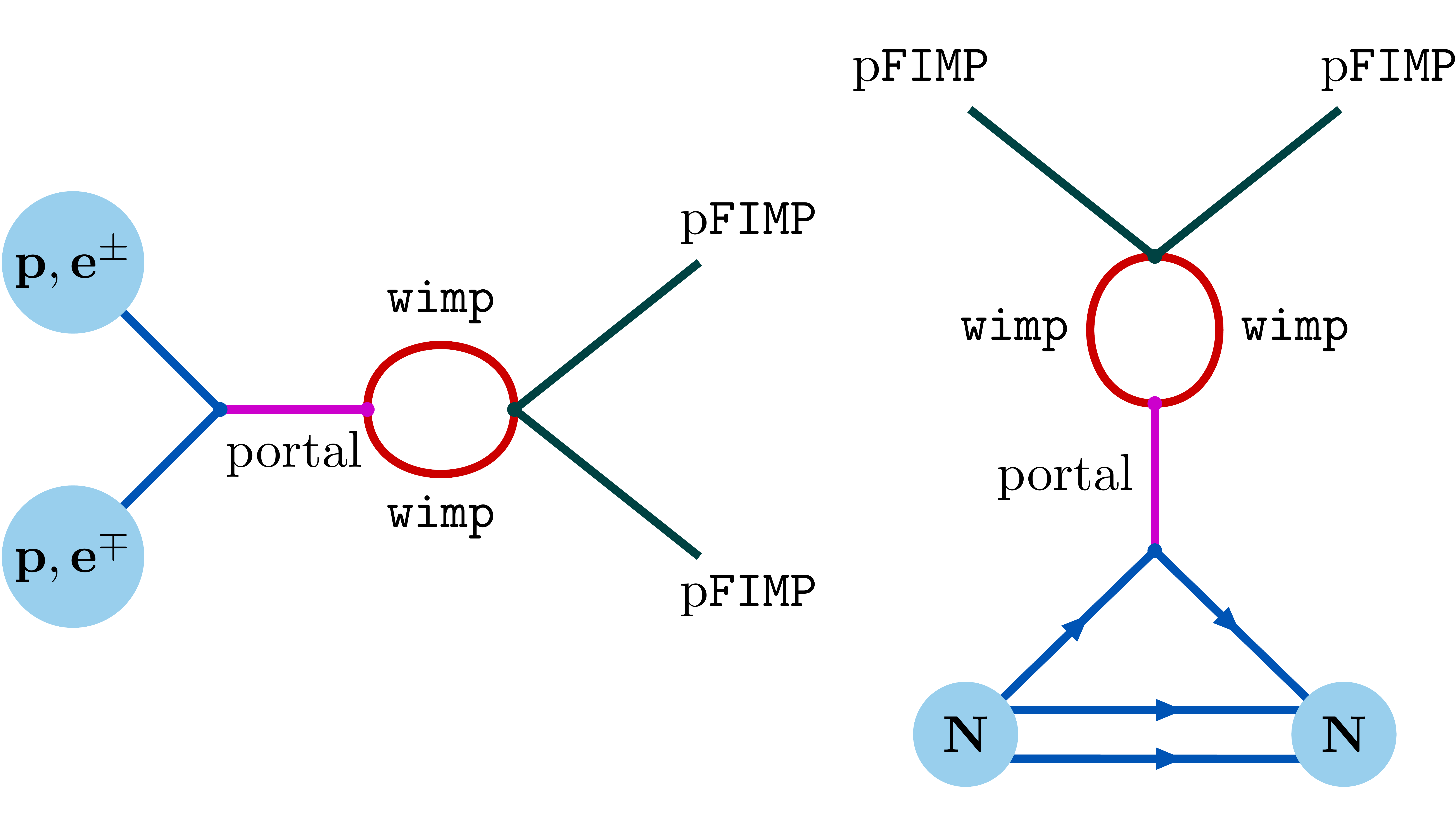}
\caption{A schematic Feynman diagram for collider (left) and direct-detection (right) of p{\tt FIMP} via one-loop {\tt WIMP} interaction.}
\label{fig:dm-collider}
\end{figure}

In Table \ref{tab:possible-scenario}, we depict a few examples of two component DM scenarios, with their strength of DM-DM conversion, possibility of producing 
correct relic abundance, along with direct (indirect) detection prospects at present/future experiments. Generically, the p{\tt FIMP} detection is heavily dependent 
on the detection prospects of the {\tt WIMP}, as the p{\tt FIMP} direct search or collider search signal proceeds via {\tt WIMP} loop as shown in Fig.~\ref{fig:dm-collider}. 
The strength of the p{\tt FIMP} interaction is less than the {\tt WIMP}, but can be of similar order when the loop factor is compensated by the large 
p{\tt FIMP-WIMP} interaction.

We would now like to validate our claims of p{\tt FIMP} characteristics for a specific model. Here we use temperature dependent $\langle\sigma v\rangle$  
in terms of the model parameters. We choose the simplest model consisting of two real scalar singlets, where $\phi_1$ behaves as {\tt WIMP} and 
$\phi_2$ as {\tt FIMP}/(p{\tt FIMP}), both of which are rendered stable under $\mathbb{Z}_2\otimes\mathbb{Z}^{\prime}_2$ symmetry. The Lagrangian density is given by,

 {\footnotesize{\bea\begin{split}
\mathcal{L}&=\mathcal{L}_{\rm SM}+\frac{1}{2}\partial_{\mu}\phi_1\partial^{\mu}\phi_1-\frac{1}{2}\mu_{\phi_1}^2\phi_1^2-\frac{1}{4!}\rm\lambda_{\phi_1}\phi_1^4-\frac{1}{2}\lambda_{ 1H}\phi_1^2H^{\dagger}H\\+&
\rm\frac{1}{2}\partial_{\mu}\phi_2\partial^{\mu}\phi_2-\frac{1}{2}\mu_{\phi_2}^2\phi_2^2-\frac{1}{4!}\lambda_{\phi_2}\phi_2^4-\frac{1}{2}\lambda_{2H}\phi_2^2H^{\dagger}H-\frac{1}{4}\lambda_{12}\phi_1^2\phi_2^2\,.
\label{Lagrangian}
\end{split}\eea}}

This model in {\tt WIMP}-{\tt WIMP} limit has been explored in many texts including \cite{Bhattacharya:2016ysw}. 
Here we focus on the p{\tt FIMP} regime, partly explored in \cite{DiazSaez:2021pfw}. For that, we choose $\phi_2$ to 
have negligible coupling with the SM, $\rm\lambda_{2H} \approx 10^{-12}$. However, in presence of the {\tt WIMP}-like $\phi_1$, 
the conversion governed by $\lambda_{12}$ plays a key role and with large $\lambda_{12}\sim1$ it reproduces all the characteristics 
of p{\tt FIMP}. Relic density as a function of $\lambda_{12}$ is plotted in the supplementary material, which shows similar features 
to that of figs.~\ref{fig:c}, \ref{fig:d} and validates the p{\tt FIMP} characteristics. The {\tt WIMP}-p{\tt FIMP} limit of this model has also been 
verified with the code {\tt micrOMEGAs4.1} \cite{Belanger:2014vza}.

Direct search of p{\tt FIMP} is possible via {\tt WIMP} loop, absent a tree-level connection to SM. The Feynman graph is shown in fig.~\ref{fig:dm-collider}, 
while the detailed calculations are provided in the supplementary material.
\begin{figure}[hptb!]
\includegraphics[width=1\linewidth]{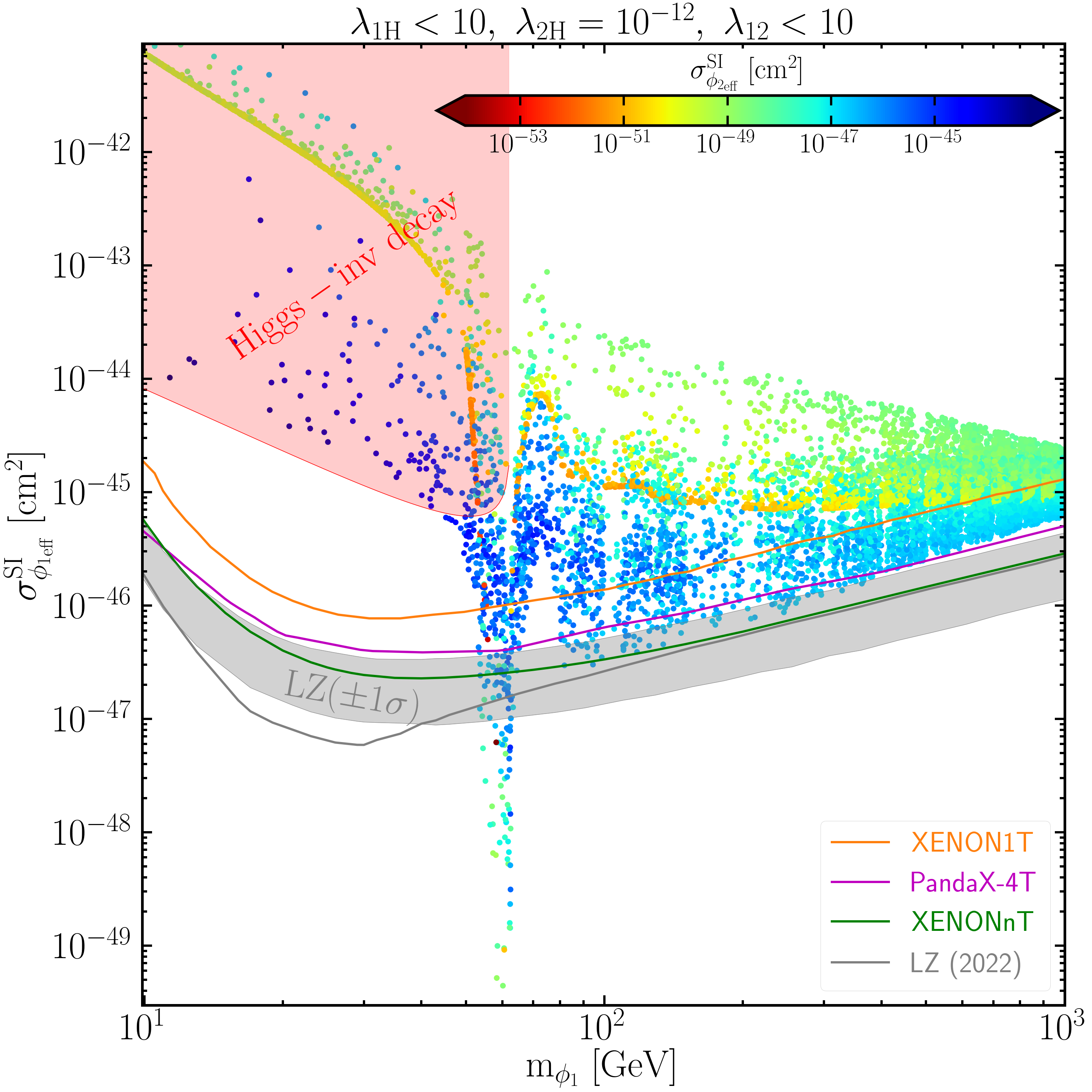}
\caption{Relic density allowed points of the model described by Eq.~\ref{Lagrangian} in $\sigma_{{\phi_1}_{\rm eff}}^{\rm SI}-m_{\phi_1}$ plane. 
The parameters varied are $\{m_{\phi_2}, \lambda_{\rm 1H}, \lambda_{12}\}$. $\sigma_{{\phi_2}_{\rm eff}}^{\rm SI}$ is shown in colour bar. 
See heading and figure inset for other parameters and experimental limits.}
\label{dd_model}
\end{figure}
\begin{table*}
\begin{ruledtabular}
\begin{tabular}{ccccccccccc}
\multirow{2}{*}{${\rm m_{\phi_1}[GeV]}$ }& \multirow{2}{*}{${\rm m_{\phi_2}[GeV]}$ }&\multirow{2}{*}{$ \lambda_{\rm 1H}$} &\multirow{2}{*}{$ \lambda_{\rm 12}$} &\multirow{2}{*}{$ \rm loop~factor$} &\multirow{2}{*}{${\rm \Omega_{\phi_1}h^2}$} &\multirow{2}{*}{${\rm \Omega_{\phi_2}h^2} $ }&\multirow{2}{*}{$\sigma^{\rm SI}_{{\phi_1}_{\rm eff}}[{\rm cm^2}]$ }&\multirow{2}{*}{$\sigma^{\rm SI}_{{\phi_2}_{\rm eff}}[{\rm cm^2}]$} &\multicolumn{2}{c}{Gluon fusion (fb)}\\\cline{10-11}
 &  & &  & & &  &  & &WIMP&pFIMP  \\\hline
  63.33&  60.50& 0.10 & 4 &$2.53\times 10^{-3}$& 0.0002&0.1203 &$3.3\times 10^{-47}$&$5.0\times 10^{-48}$ & 8.4  & 13.9 \\
 104.1&  100.0&  0.22&6  &$8.40\times 10^{-3}$ &0.0002 &0.1210&$7.89\times 10^{-47}$   &$2.15\times 10^{-47}$ & 0.69  & 1.18 $\times 10^{-3}$  \\
 405.0 &  402.9&  0.30& 8 & $1.52\times 10^{-2}$ &0.0101 &0.1091&$4.06\times 10^{-46}$   &$7.49\times 10^{-48}$ & 9$\times 10^{-4}$  & 2.34 $\times 10^{-6}$ \\
\end{tabular}
\end{ruledtabular}
\caption{Some benchmarks for the two-component scalar singlet model (Eq.~\eqref{Lagrangian}) 
in {\tt WIMP}-p{\tt FIMP} limit, where the loop factor is approximated as $\sim(\lambda_{\rm 1H}\lambda_{12})/(16\pi^2)$. 
The last two columns depict the {\tt WIMP} and p{\tt FIMP} production cross-section at LHC via Gluon fusion yielding mono-jet+$\slashed{E_T}$ in the final state at $\sqrt{s}=14~\rm TeV$.}
\label{tab:bench-mark}
\end{table*}
Large {\tt WIMP}-p{\tt FIMP} interaction ($\lambda_{12}$) helps in the detection of p{\tt FIMP}, 
the loop suppression factor $\sim 1/(16\pi^2)$ allows it to evade existing bound and make it accessible to future sensitivities.
In fig.~\ref{dd_model}, spin-independent (SI) direct detection cross-section for the {\tt WIMP} ($\phi_1$) and p{\tt FIMP} $(\phi_2)$ 
is shown as a function of {\tt WIMP} mass for relic density allowed points of the model. The sensitivity of existing direct search data 
have all been compared~\cite{XENON:2018voc,XENON:2023sxq,PandaX-4T:2021bab,LZ:2022ufs}. 
A clear dip is seen in the resonance region $m_{\phi_1}\sim m_h/2$, where due to resonance enhancement $\lambda_{\rm 1 H}$ 
is relaxed and points satisfy direct search constraints. One should however remember the limitations of Lee-Weinberg 
mechanism in the vicinity of resonance as pointed out by \cite{Binder:2017rgn,Ala-Mattinen:2019mpa}.

When $m_{\phi_1} > m_{\phi_2}$, small $\delta m$ enhances the $\phi_1\phi_1 \to \phi_2\phi_2$ conversion, 
consequently decreasing $\Omega_{\phi_1}$ and effective direct detection cross-section 
$\sigma^{\rm SI}_{i_{\rm eff}}=\frac{\Omega_i}{\Omega_{\rm tot}}\sigma^{\rm SI}_i$, where $i=\{\phi_1,\phi_2\}$. 
However, most of the parameter space of this {\tt WIMP}-p{\tt FIMP} limit of the two component scalar DM model 
is on the verge of exclusion by the LZ (2022) data, leaving the Higgs resonance, $m_{\phi_1}\sim m_h$ regions and some points close to the $\pm 1\sigma$ exclusion limit up to 500 GeV survive to be explored. 
Unlike {\tt WIMP}-{\tt WIMP} case, where two DMs may have widely different masses, p{\tt FIMP} limit necessitates the mass splitting to be 
small and therefore serves as a guiding principle for searching both DM components. Similarly p{\tt FIMP}s can also be produced in collider 
via one-loop {\tt WIMP} graphs yielding mono-X plus missing energy signals as estimated in the Table \ref{tab:bench-mark} for some benchmark points 
where the model satisfies relic density and direct search bounds. In this case, detection of p{\tt FIMP} seems only possible in the resonance region, 
owing to a resonant enhancement in the production. Interesting to note that, even in this region, the effective coupling between Higgs and p{\tt FIMP} 
is rather small due to loop suppression, thereby making it safe from the Higgs invisible decay constraints. One can have plethora of other possibilities, 
like a fermion {\tt WIMP} instead of a scalar {\tt WIMP}, which enhances the detection possibility of a scalar p{\tt FIMP}, 
see for example, \cite{Bhattacharya:2022vxm}. We again note here that the model chosen for illustration is not the best for detection, rather 
it is the simplest one.

Furthermore, p{\tt FIMP} having a similar mass to that of the {\tt WIMP} provides a challenge in distinguishing 
the DM components in direct and collider searches, as the distinguishability (a kink in the recoil spectrum for direct search 
\cite{Herrero-Garcia:2017vrl,Herrero-Garcia:2018qnz} or two bumps in missing energy at colliders \cite{Bhattacharya:2022wtr,Bhattacharya:2022qck}) 
often depends on the mass splitting of the DM components. However mono-X signal can be useful, 
as the position of the peak here primarily depends on the operator structure, angular momentum conservation etc. 
We are exploring such possibilities, however the two component scalar case will be difficult to disentangle.

Finally we summarise here. We have shown that the {\tt FIMP} can reach equilibrium
in presence of a thermal DM component and sizeable conversion cross-section. We call it p{\tt FIMP}.
The freeze-out of p{\tt FIMP} precedes that of its thermal partner and requires a small mass splitting with it. The numerical solution of Eq.~\ref{eq:cbeq}
presented here and approximate analytical solution provided in the supplementary material validates all the p{\tt FIMP} properties 
in a model independent way. The p{\tt FIMP} realization in presence of {\tt SIMP} is also provided in the supplementary material. The freeze-out properties of  
p{\tt FIMP} is verified for the simplest two-component DM model with two scalar singlets. The detection prospect of p{\tt FIMP} via {\tt WIMP} loop in future 
direct, indirect and collider search experiments opens up new possibilities. Such investigations in context of direct DM search for two-component {\tt WIMP-}p{\tt FIMP} 
model has been pursued in \cite{Bhattacharya:2022vxm}.

{\bf Acknowledgments:---} SB and JL acknowledge the grant CRG/2019/004078 from SERB, Govt. of India. DP thanks University Grants Commission for senior research fellowship and Heptagon, IITG for useful discussions.

\newpage
\appendix
\onecolumngrid
\begin{center}
 {\bf \large SUPPLEMENTAL MATERIAL}
\end{center}
\section{Semi-analytic solution of the cBEQ for {\tt {WIMP}}-p{\tt FIMP} scenario}
\label{analytic}
A two-component {\tt {WIMP}}-p{\tt FIMP} framework is governed by a coupled Boltzmann equations (cBEQ),
{\footnotesize\begin{center}\begin{eqnarray}\begin{split}
 \frac{dY_1}{dx}=-\frac{2\pi^2\rm{M_{pl}}}{45\times 1.67}\frac{g_{\star}^{ s}}{ \sqrt{g_{\star}^{ \rho}}}\frac{\mu_{12}}{x^2}&\Biggl[\langle\sigma v\rangle_{11\to\rm{SM}}\Bigl(Y_1^2-Y_{1}^{\rm{eq}^2}\Bigr)+\langle\sigma v\rangle_{11\to22}\Bigl(Y_1^2-\frac{Y_1^{\rm eq^2}}{Y_2^{\rm eq^2}}Y_2^2\Bigr)\Biggr],\\
\frac{dY_2}{dx}=\frac{2\rm{M_{pl}}}{1.67\times \sqrt{g_{\star}^{\rho}}}\frac{x}{\mu_{12}^2}\langle\Gamma_{\rm SM\to22}\rangle (Y_{\rm SM}^{\rm eq }-\frac{Y_2^2}{Y_2^{\rm eq^2}}Y_{\rm SM}^{\rm eq})+&\frac{4\pi^2\rm{M_{pl}}}{45\times 1.67}\frac{g_{\star}^{ s}}{ \sqrt{g_{\star }^{\rho}}}\frac{\mu_{12}}{x^2}\Biggl[\langle\sigma v\rangle_{\rm{SM}\to 22}\Bigl(Y_{\rm SM}^{\rm eq^2}-\frac{Y_2^2}{Y_2^{\rm eq^2}}Y_{\rm SM}^{\rm eq^2}\Bigr)+\langle\sigma v\rangle_{11\to22}\Bigl(Y_1^2-\frac{Y_{1}^{\rm eq^2}}{Y_{2}^{\rm eq^2}}Y_2^2\Bigr)\Biggr]\,.\end{split}\label{eq:cbeq}
\end{eqnarray}\end{center}}
We elaborate upon the semi analytical solution of the cBEQ as in Eq.~\ref{eq:cbeq} in {\tt WIMP-pFIMP} limit, for both the hierarchies:
\begin{itemize}
\item{} Case-I: {\tt WIMP} mass $>$ p{\tt FIMP} mass (i.e. $m_{1}>m_{2}$).
\item{} Case-II: {\tt WIMP} mass $<$ p{\tt FIMP} mass (i.e. $m_{1}<m_{2}$).
\end{itemize}
The methodology follows the procedure in \cite{Bhattacharya:2013hva}.
\subsection{Case-I ($m_{1}>m_{2}$)}
In {\tt WIMP-pFIMP} limit, we neglect the p{\tt FIMP} production from the visible sector as the p{\tt FIMP} coupling with SM is very small, compared to the conversion cross-section. 
So, in the following we consider, $\Gamma_{\rm SM\to22}=\langle\sigma v\rangle_{\rm{SM}\to 22}=0$. For notational ease, let us also use, 
$\rm\langle\sigma v\rangle_{11\to SM}=\sigma_1,~\langle\sigma v\rangle_{11\to 22}=\sigma_{12},~\langle\sigma v\rangle_{22 \to 11}=\sigma_{21}$, 
$\rm [(2\pi^2 \sqrt{g_{\star}^{\rho}})/(45\times 1.66)]M_{pl}=n_{f}$ with Planck mass $\rm M_{pl}=1.22 \times 10^{19}$ GeV. Recall that 
$\sigma_{12}=\sigma_{21}\left(n_2^{eq}/{n_{1}^{eq}}\right)^2$.  With above, the cBEQ in terms of bath temperature $(T)$ becomes,
\bea\begin{split}
 &\frac{dY_1}{dT}=n_{f}\left[\sigma_1\left(Y_1^2-Y_1^{\rm eq^2}\right)+\sigma_{12}\left(Y_1^2-\frac{Y_1^{\rm eq^2}}{Y_2^{\rm eq^2}}Y_2^2\right)\right]\,,\\
&\frac{dY_2}{dT}=-2n_{f}\left[\sigma_{12}\left(Y_1^2-\frac{Y_{1}^{\text{eq}^2}}{Y_{2}^{\text{eq}^2}}Y_2^2\right)\right]\,.
\end{split}\label{cbeq}
\eea
The equilibrium yield is given by Maxwell-Boltzmann distribution, 
\bea
Y_{i}^{\rm eq}\left(T\right) = 0.145\frac{g}{g_{\star}^s}\left(\frac{m_i}{T}\right)^{3/2}\exp\left(-\frac{m_i}{T}\right)=A_iT^{-3/2}e^{-m_i /T}~\text{with}~A_i=0.145\frac{g}{g_{\star}^s}m_i^{3/2}\,. \nonumber
\eea
Defining $Y_{1}^{eq^2}\left(\frac{\sigma_1}{\sigma_1+\sigma_{12}}+\frac{\sigma_{12}}{\sigma_1+\sigma_{12}}\frac{Y_2^2}{Y_2^{eq^2}}\right)=y_1^{eq^2}$, \autoref{cbeq} becomes,
\bea\begin{split}
 &\frac{dY_1}{dT}=n_{f}\left(\sigma_1+\sigma_{12}\right)\left(Y_1^2-y_1^{\rm eq^2}\right)\,,\\
&\frac{dY_2}{dT}=-2n_{f}\left[\sigma_{12}\left(Y_1^2-\frac{Y_1^{\text{eq}^2}}{Y_{2}^{\text{eq}^2}}Y_2^2\right)\right]\,.
\end{split}\label{modi-cbeq}
\eea
In order to solve the cBEQ analytically, it would be convenient to divide this whole scenario in three regions on the bath temperature: 
\textbf{Region A}: $T\gg T_{f_i}$, \textbf{Region B}: $T\simeq T_{f_i}$ and \textbf{Region C}: $T\ll  T_{f_i}$ where {\color{blue}$T_{f_i}$} denotes 
the freeze-out temperature of both species $ i=1,2$. 
Let use consider the difference of DM yield from equilibrium yield as $\Delta_1=Y_1-y_1^{\rm eq}$ and $\Delta_2=Y_2-Y_2^{\rm eq}$. Using these relations, \autoref{modi-cbeq} becomes,
\bea\begin{split}
 &\frac{d\Delta_1}{dT}+ \frac{dy_{1}^{eq}}{dT}=n_{f}\left(\sigma_1+\sigma_{12}\right)\left(\Delta_1^2+2\Delta_1y_1^{\rm eq}\right)\,,\\
& \frac{d\Delta_2}{dT}+ \frac{dY_{2}^{eq}}{dT}=-2n_{f}\sigma_{12}\Bigl[\left(\Delta_1+y_1^{\rm eq}\right)^2-\frac{Y_1^{\rm eq^2}}{Y_2^{\rm eq^2}}\left(\Delta_2+Y_2^{\rm eq}\right)^2\Bigr]\,.
\end{split}\label{cbeq1a}
\eea

\subparagraph{$\bullet$ \textbf{Region A:\\}}

For $T\gg T_{f_i}$, $\frac{d\Delta_i}{dT}$ is negligible and \autoref{cbeq1a} becomes,
\begin{align}
 & \frac{dy_{1}^{eq}}{dT}=n_{f}\left(\sigma_1+\sigma_{12}\right)\left(\Delta_1^2+2\Delta_1y_1^{\rm eq}\right)\,,\label{cbeqd}\\
& \frac{dY_{2}^{eq}}{dT}=-2n_{f}\sigma_{12}\Bigl[\left(\Delta_1+y_1^{\rm eq}\right)^2-\frac{Y_1^{\rm eq^2}}{Y_2^{\rm eq^2}}\left(\Delta_2+Y_2^{\rm eq}\right)^2\Bigr]\,.\label{cbeqe}
\end{align}
\subparagraph{ $\bullet$ \textbf{Region B:\\}}
Let us suppose $T_{f_2}>T_{f_1}$, then both \autoref{cbeqd} and \autoref{cbeqe} hold at $T_{f_2}$, but only \autoref{cbeqd} holds at $T_{f_1}$ as p{\tt FIMP} already freezes 
out at $T_{f_1}$. In the vicinity of $T\simeq T_{f}$, we may further assume $\Delta_1(T_{f_1})=c_1y_{ 1}^{eq}(T_{f_1})$ and $\Delta_2(T_{f_2})=c_2Y_{ 2}^{eq}(T_{f_2})$~\cite{Kolb:1990vq}, 
where $c_i$'s are unknown constants whose values are determined by matching the analytical result with the numerical one. Substituting in \autoref{cbeqd} and  \autoref{cbeqe}, we get,
\begin{eqnarray}
&y_{1}^{eq^{\prime}}(T_{f_1})\approx n_{f}\left(\sigma_1+\sigma_{12}\right)c_1(c_1+2)y_{1}^{eq^2}(T_{f_1}) \label{cbeqf}\,,\\
& 2n_{f}(\sigma_1+\sigma_{12})\sigma_{12}\left[\left(c_2+1\right)^2Y_{1}^{eq^2}(T_{f_2})-y_{1}^{eq^2}(T_{f_2})\right]\approx (\sigma_1+\sigma_{12})Y_{2}^{eq^{\prime}}(T_{f_2})+2\sigma_{12}y_1^{eq^{\prime}}(T_{f_2})\,.\label{cbeqg}
\end{eqnarray}
In the above,
{\small
\begin{align}
Y_i^{\rm eq^{\prime}}(T)&=\left(-\frac{3}{2T}+\frac{m_i}{T^2}\right)Y_i^{\rm eq}(T)\,,\\
y_1^{eq^2}(T_{f_2})&=Y_{1}^{eq^2}(T_{f_2})\left(\frac{\sigma_1}{\sigma_1+\sigma_{12}}+\frac{\sigma_{12}}{\sigma_1+\sigma_{12}}\left(c_2+1\right)^2\right)\,,\\
y_1^{eq^2}(T_{f_1})&=Y_{1}^{eq^2}(T_{f_1})\left(\frac{\sigma_1}{\sigma_1+\sigma_{12}}+(c_2+1)^2\frac{\sigma_{12}}{\sigma_1+\sigma_{12}}\frac{Y_2^{eq^2}(T_{f_2})}{Y_2^{eq^2}(T_{f_1})}\right)\,,\\
y_1^{eq^{\prime}}(T_{f_2})&=\frac{Y_{1}^{eq^2}(T_{f_2})}{y_{1}^{eq}(T_{f_2})}\left[\left(-\frac{3}{2T_{f_2}}+\frac{m_1}{T_{f_2}^2}\right)\left(\frac{\sigma_{1}}{\sigma_{1}+\sigma_{12}}+\frac{\sigma_{12}}{\sigma_{1}+\sigma_{12}}\left(c_2+1\right)^2\right)-\left(-\frac{3}{2T_{f_2}}+\frac{m_2}{T_{f_2}^2}\right)\frac{\sigma_{12}}{\sigma_{1}+\sigma_{12}}c_2\left(c_2+1\right)\right]\,,\\
y_1^{eq^{\prime}}(T_{f_1})&=\frac{Y_{1}^{eq^2}(T_{f_1})}{y_{1}^{eq}(T_{f_1})}\left[\left(-\frac{3}{2T_{f_1}}+\frac{m_1}{T_{f_1}^2}\right)\frac{\sigma_{1}}{\sigma_{1}+\sigma_{12}}+\frac{\sigma_{12}}{\sigma_{1}+\sigma_{12}}\left(c_2+1\right)^2\frac{m_1-m_2}{T_{f_1}^2}\left(\frac{T_{f_1}}{T_{f_2}}\right)^3e^{-2\frac{m_2}{T_{f_2}}}e^{2\frac{m_2}{T_{f_1}}}\right]\,.
\end{align}}
\autoref{cbeqf} and \autoref{cbeqg} can be written further as,
\begin{eqnarray}
&y_{1}^{eq^{\prime}}(T_{f_1})=n_{f}\left(\sigma_1+\sigma_{12}\right)c_1(c_1+2)y_{1}^{eq^2}(T_{f_1})\,,
\label{cbeqi}\\
&P ~T_{f_2}^{-1/2}=2\sigma_{12}QT_{f_2}e^{\frac{m_1}{T_{f_2}}}+(\sigma_{1}+\sigma_{12})\left(-\frac{3}{2}+\frac{m_2}{T_{f_2}}\right)\left(\frac{m_2}{m_1}\right)^{3/2}e^{2\frac{m_1}{T_{f_2}}}e^{-\frac{m_2}{T_{f_2}}}\,;\label{cbeqh}
\end{eqnarray}
where,
{\small
\begin{eqnarray}
&P=2A_1n_{f}\sigma_{1}\sigma_{12}c_2\left(c_2+2\right)\,,\\&
Q=\left[\left(-\frac{3}{2T_{f_2}}+\frac{m_1}{T_{f_2}^2}\right)\left(\frac{\sigma_{1}}{\sigma_{1}+\sigma_{12}}+\frac{\sigma_{12}}{\sigma_{1}+\sigma_{12}}\left(c_2+1\right)^2\right)-\left(-\frac{3}{2T_{f_2}}+\frac{m_2}{T_{f_2}^2}\right)\frac{\sigma_{12}}{\sigma_{1}+\sigma_{12}}c_2\left(c_2+1\right)\right]\left[\frac{\sigma_1}{\sigma_1+\sigma_{12}}+\frac{\sigma_{12}}{\sigma_1+\sigma_{12}}\left(c_2+1\right)^2\right]^{-\dfrac{1}{2}}\,.
\end{eqnarray}}
\autoref{cbeqi} and \autoref{cbeqh} are transcendental equations, analytical solutions are difficult to obtain, 
but using numerical method it is possible to extract the values of freeze out temperature $T_{f_1}$ and $T_{f_2}$, which we did.
\subparagraph{$\bullet$ \textbf{Region C:\\}}
When $T\ll T_{f_i},$ then, $Y_i^{\rm eq},~Y_i^{\rm eq^{\prime}}$ and $\frac{Y_{1}^{eq^2}}{Y_{2}^{eq^2}}$ are exponentially suppressed, so, $\Delta_i\gg Y_i^{\rm eq}$ and \autoref{cbeq1a} becomes,
\begin{align}
&\frac{d\Delta_1}{dT}\approx n_{f}(\sigma_1+\sigma_{12})\Delta_1^2\label{cbeql}\,,\\
 &\frac{d\Delta_2}{dT}\approx-2n_{f}\sigma_{12}\Delta_1^2\label{cbeqm}\,.
\end{align}
After solving the differential equations between $T_{f_i}$ to $T$ with $\Delta_i(T)\gg\Delta_i(T_{f_i})$, we get,
{\small
\begin{eqnarray}
&\Delta_1(T)\approx\left[\frac{1}{\Delta_1(T_{f_1})}+n_{f}(\sigma_1+\sigma_{12})\left(T_{f_1}-T\right)\right]^{-1}\,,\label{cbeqn}\\
&\Delta_2(T)\approx\Delta_2(T_{f_2})-2n_{f}\sigma_{12}(T-T_{f_2})\left[\left(\frac{1}{\Delta_1(T_{f_1})}+n_{f}(\sigma_1+\sigma_{12})(T_{f_1}-T)\right)\left(\frac{1}{\Delta_1(T_{f_1})}+n_{f}(\sigma_1+\sigma_{12})(T_{f_1}-T_{f_2})\right)\right]^{-1}\,.
\label{cbeqo}
\end{eqnarray}}
Following, 
$\Delta_1(T_{f_1})\approx c_1 y_1^{\rm eq^{}}(T_{f_1})$, $\Delta_2(T_{f_2})\approx c_2 Y_2^{\rm eq^{}}(T_{f_2})$, at $T=T_{\rm CMB}=2.35\times 10^{-13}\rm GeV$, the DM yields are,
 \begin{align}
&\Delta_1(T_{\rm CMB})\approx Y_1(T_{\rm CMB})\approx\left(\frac{1}{\Delta_1(T_{f_1})}+n_{f}(T_{f_1}-T_{\rm CMB})(\sigma_1+\sigma_{12})\right)^{-1}\,,\\
&\Delta_2(T_{\rm CMB})\approx Y_2(T_{\rm CMB})\approx\Delta_2(T_{f_2})+2n_{f}\sigma_{12}\int_{T_{\rm CMB}}^{T_{f_2}}\Delta_1^2(T) ~dT\,.
\end{align}
\begin{figure*}[hptb!]
\includegraphics[width=0.45\linewidth]{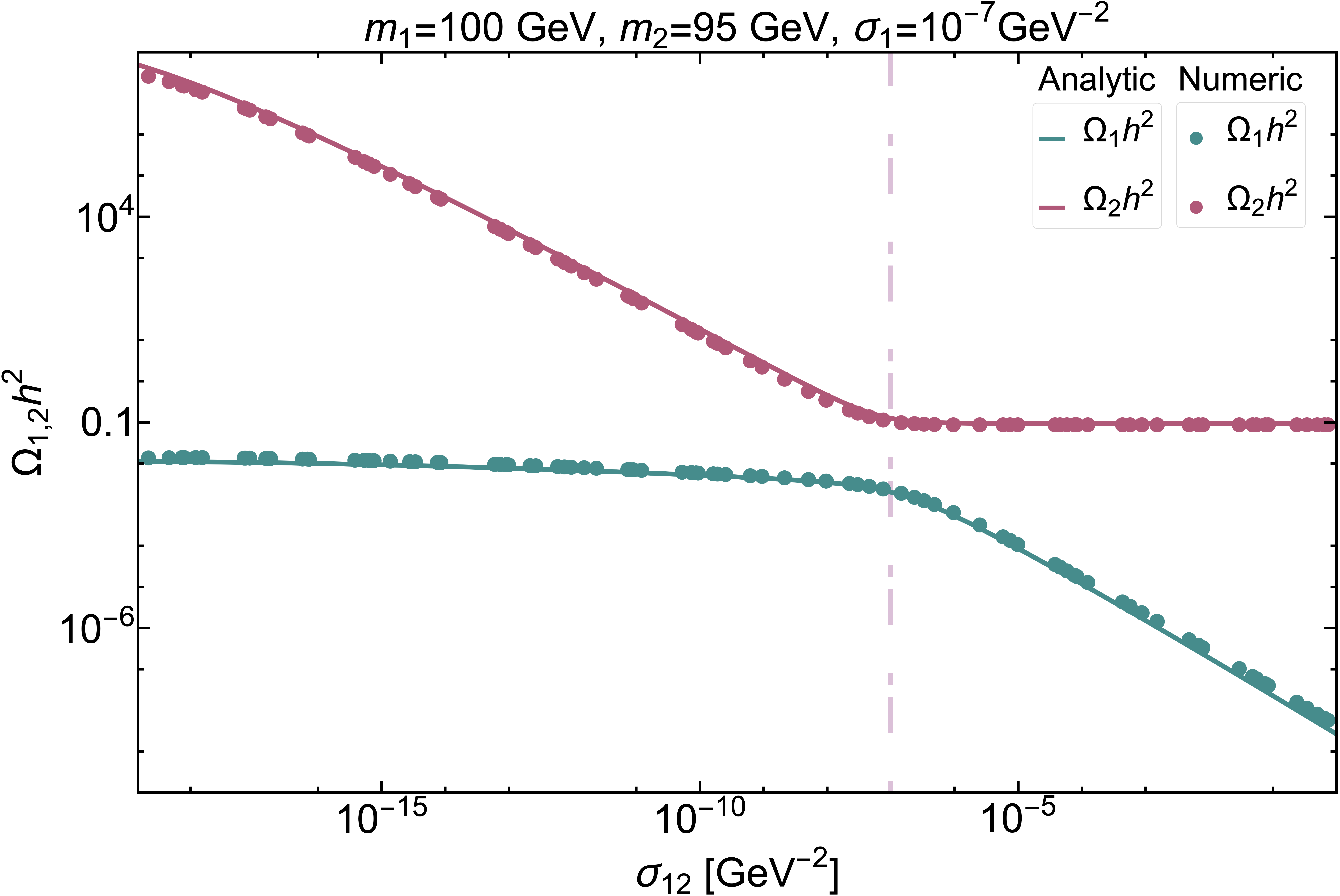}
\caption{Analytic solution of the cBEQ (represented by thick line) and numerical solution (represented by points) of Eq.~\eqref{eq:cbeq} for p{\tt FIMP-WIMP} case
is shown as a function of conversion cross-section. The unknown constants are chosen as $c_{1}=2$ and $c_2=7$. Vertical dot-dashed line corresponding to $\sigma_{12}=\sigma_1$.}
\label{fig:mi-analytic}
\end{figure*}
\begin{figure*}[htb!]
\includegraphics[width=0.33\linewidth]{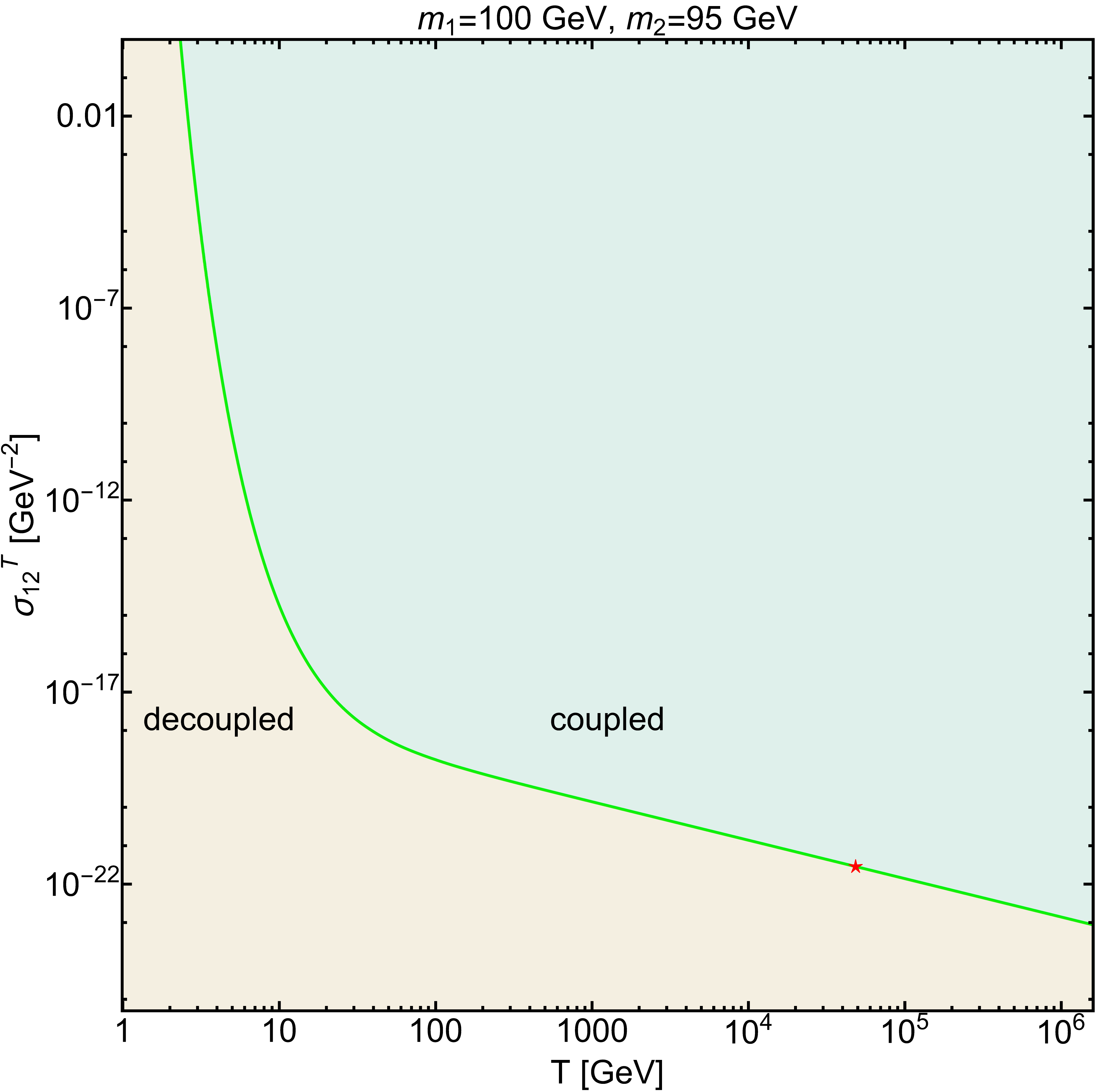}~~
\includegraphics[width=0.33\linewidth]{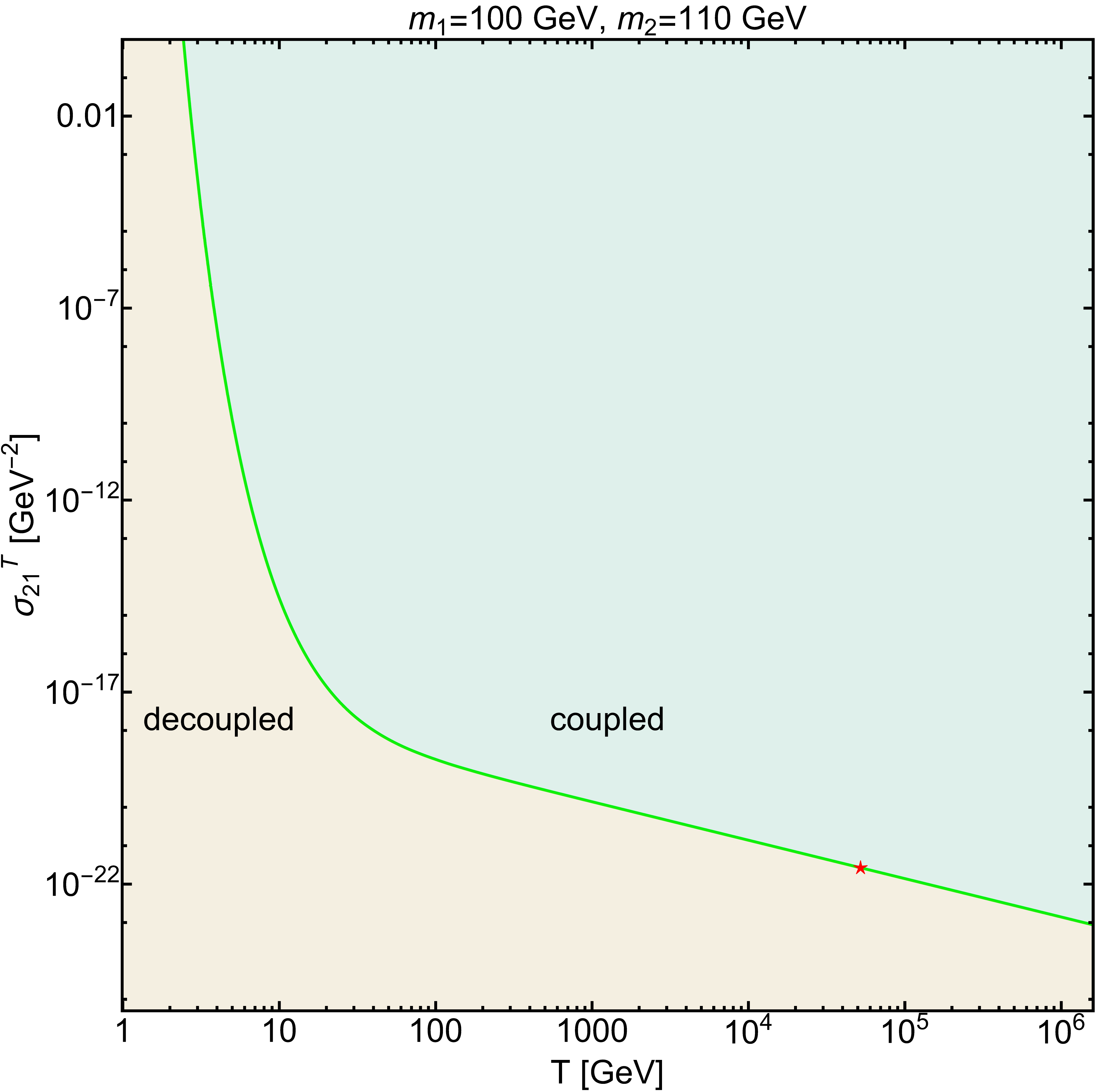}
\caption{In-equilibrium $\rm (\Gamma>H)$ (green shade) and out-of-equilibrium $\rm(\Gamma<H)$ (yellow shade) possibilities for p{\tt FIMP}. 
Green line corresponds to $\rm \Gamma(T)=H(T)$ and red star corresponds to $\rm T=T_{RH}$. Both mass hierarchies $m_{1}>m_{2}$ (left), $m_{2}>m_{1}$ (right) are plotted.}
\label{couple_decouple}
\end{figure*}
Using the analytical approximate solutions, relic density as a function of conversion cross-section for both {\tt WIMP} and p{\tt FIMP} 
have been shown in \autoref{fig:mi-analytic}. They show reasonably good agreement, the tiny mismatch with numeric result at low and high 
conversion region is only due to the entropy d.o.f $(g_*^s)$ and matter d.o.f $(g_*^{\rho})$ which depends on temperature (specifically $T\lesssim5\rm ~GeV$) 
but for simplicity we have neglected this temperature dependence and chosen a fixed value of $g_*^{\rho}= g_*^s \simeq 75$ for the semi analytical solution.  

In the similar way it is easy to evaluate the present DM relic for the inverse mass hierarchy ($m_{1}<m_{2}$) where one necessary input is $T_{f_1}>T_{f_2}$ and 
$\Delta_2(T_{f_2})=c_2y_2^{eq}(T_{f_2})$ with $y_2^{ eq}=Y_1\frac{Y_2^{ eq}}{Y_1^{ eq}}$. 
The transition of {\tt FIMP} into a p{\tt FIMP} depends not only on the temperature but also on the interaction rate. This can be understood, just by analysing the 
relation between interaction rate ($\Gamma$) and Hubble expansion rate ($H$). \autoref{couple_decouple} depicts it for both mass hierarchies: 
$m_{1}>m_{2}$ (left), $m_{2}>m_{1}$ (right). $\rm\Gamma\simeq H$ has been represented by the green curve. 
The red star points correspond to ${\rm\Gamma(T\sim T_{RH})=H(T\sim T_{RH})}$ in both the figures. In our model-independent analysis, 
we have neglected the temperature dependence of the conversion cross-section, which may alter the high temperature behaviour of the plot.

\section{Dynamics of the p{\tt FIMP} in presence of a {\tt SIMP}}\label{pFIMP-simp}
\begin{figure*}[hptb!]
\subfloat[]{\includegraphics[width=0.475\linewidth]{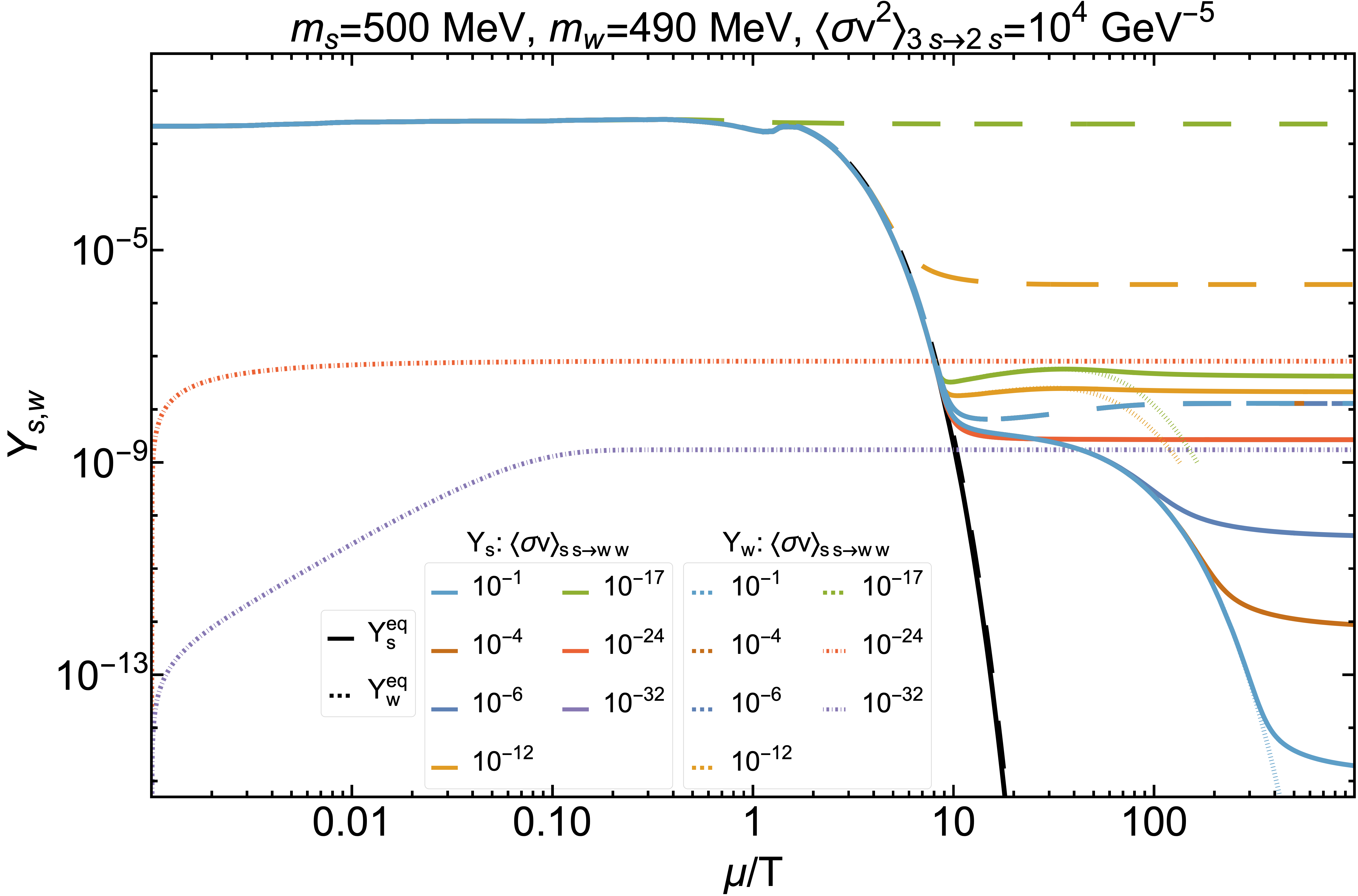}}~~
\subfloat[]{\includegraphics[width=0.475\linewidth]{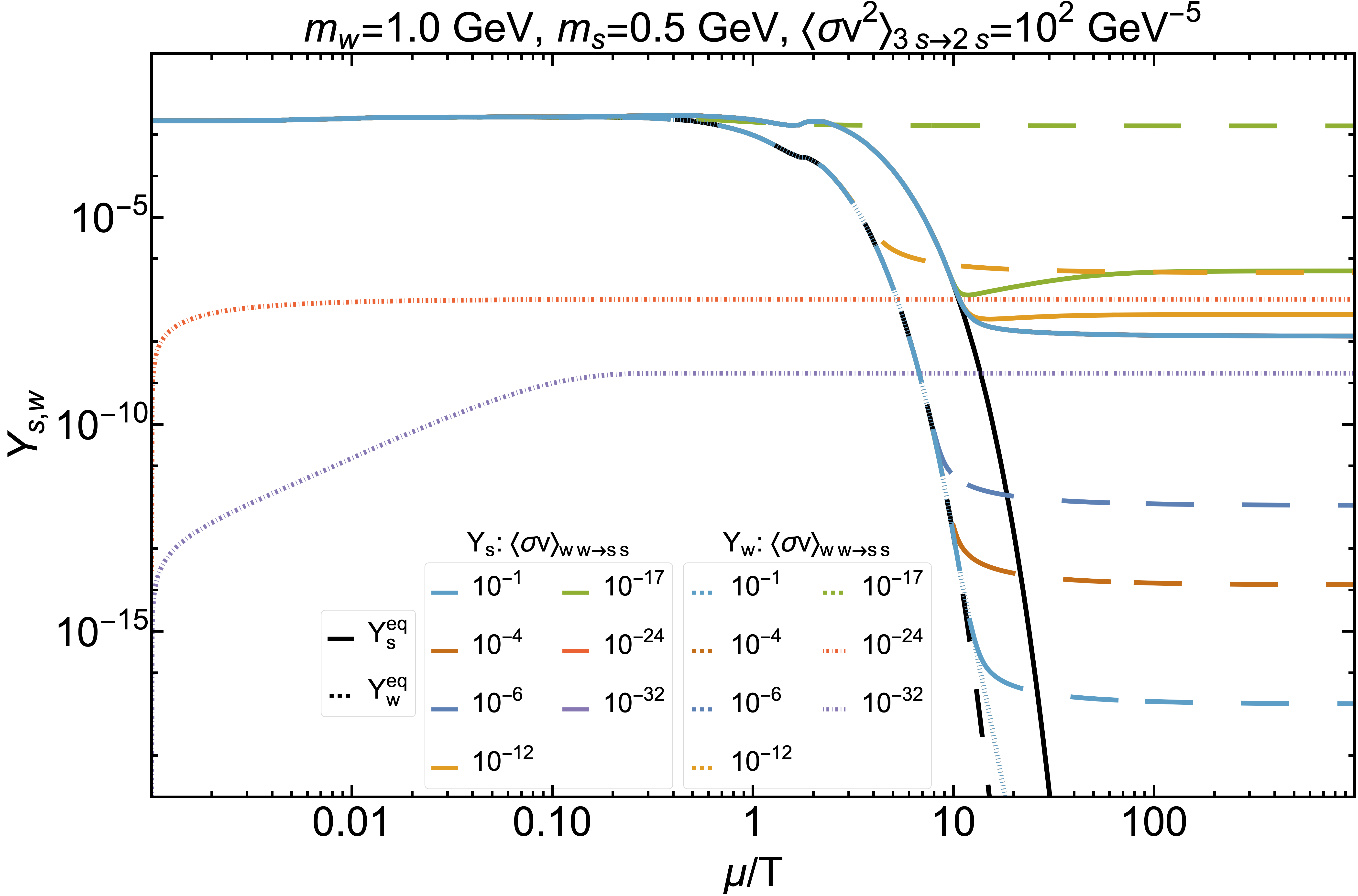}}\\
\caption{DM yield with $x=\mu/T$ in p{\tt FIMP-SIMP} case for fixed DM masses and conversion cross-sections, where the black solid (dashed) curves denote 
equilibrium distribution for {\tt SIMP} (p{\tt FIMP}) corresponding to (a) $m_{\rm s}>m_{\rm w}$ and (b) $m_{\rm s}<m_{\rm w}$. For all the plots 
$\langle \sigma v\rangle_{\rm s~ s\to  SM~SM}\,=\, 10^{-10}\rm GeV^{-2},$  $\rm\langle\Gamma\rangle_{SM\to\rm w~w}\,=\,10^{-23}\rm GeV^{-1},$ 
$\langle \sigma v\rangle_{\rm SM~SM\to \rm w~w}\,=\,10^{-32}\rm GeV^{-2}$ $\rm~and ~m_{\rm SM}\,=\,10 ~ GeV$ have been assumed. 
}\label{fig:simp-pFIMP}
\end{figure*}

The p{\tt FIMP} behaviour could also be achieved in presence of {\tt SIMP} \cite{Hochberg:2014dra} when DM-DM conversion is sufficiently large 
($\sim$ weak scale). The cBEQ for p{\tt FIMP}-{\tt SIMP} scenario can be written as:
{\footnotesize
\begin{align}
\frac{dY_{\rm w}}{dx}=&\frac{2~\bf{s}}{x~H(x)}\left[\frac{1}{\textbf{s}} \left(Y_{\rm SM}^{\rm eq }-Y_{\rm SM}^{\rm eq}\frac{Y_{\rm w}^2}{Y_{\rm w}^{\rm eq^2}}\right)\langle\Gamma\rangle_{\rm SM\to\rm w~w}\,+\,\Bigl(Y_{\rm SM}^{\rm eq^2}-Y_{\rm SM}^{\rm eq^2}\frac{Y_{\rm w}^2}{Y_{\rm w}^{\rm eq^2}}\Bigr)\langle\sigma v\rangle_{\rm{SM~ SM}\to\rm w ~w}+\Bigl(Y_{\rm s}^2-Y_{\rm s}^{\rm eq^2}\frac{Y_{\rm w}^2}{Y_{\rm w}^{\rm eq^2}}\Bigr)\langle\sigma v\rangle_{\rm s~s\to w~w}\right]\,,\\
\frac{dY_{\rm s}}{d x}=&-\frac{\bf{s}}{x~H(x)}\bigg[\left(Y_{\rm s}^2-Y_{\rm s}^{\rm eq^2}\right)\langle\sigma v\rangle_{\rm s~s\to SM~SM}\,+\,\textbf{s}\left(Y_{\rm s}^3-Y_{\rm s}^2Y_{\rm s}^{\rm eq}\right)\langle\sigma v^2\rangle_{3s\to 2s}+\left(Y_{\rm s}^2-Y_{\rm s}^{\rm eq^2}\frac{Y_{\rm w}^2}{Y_{\rm w}^{\rm eq^2}}\right)\langle\sigma v\rangle_{\rm s~s\to w~w}\bigg]\,.
\end{align}}
In the above subscripts $s$ denote {\tt SIMP} and ${\rm w}$ denote p{\tt FIMP}, 
$\mu=1/\left(\frac{1}{m_{\rm w}}+\frac{1}{m_{\rm s}}\right)^{-1},~x=\mu/T,~H(x)\,=\,1.67\sqrt{g_*^{\rho}}\mu^2x^{-2}\rm M_{pl}^{-1},~$$\bf{s}$$\,=\,\frac{ 2\pi^2}{45}g_{*}^{\rm \bf{s}}\left(\frac{\mu}{x}\right)^3$$,~\langle\sigma v\rangle_{\rm w~w\to s~s}\,=\,\langle\sigma v\rangle_{\rm s~s\to w~w}\left(Y_{\rm s}^{\rm eq}/Y_{\rm w}^{\rm eq}\right)^2,~Y_{ i}^{\rm eq}\,=\,\frac{45}{4\pi^4}\frac{g_i}{g_*^{\bf s}}\left(\frac{m_i}{\mu}x\right)^2K_2\left(\frac{m_i}{\mu}x\right)$. For simplicity we only have taken $3{\rm DM\to 2DM}$ depletion process for {\tt SIMP}, but 
${\rm(nDM\to 2DM}$ with $n>3)$ can also be taken to show the same effect. We solve the cBEQ with increasing order of conversion cross-section to show the 
change from {\tt FIMP} to p{\tt FIMP} case in \autoref{fig:simp-pFIMP} for both mass hierarchies $m_{\rm w}>m_{\rm s}$ (left panel) and $m_{\rm w}<m_{\rm s}$ (right panel). 
The features remain very similar to p{\tt FIMP}-{\tt WIMP} scenario. 
\section{Solution to cBEQ for two component scalar DM model}
\begin{figure*}[hptb!]
\subfloat[]{\includegraphics[width=0.45\linewidth]{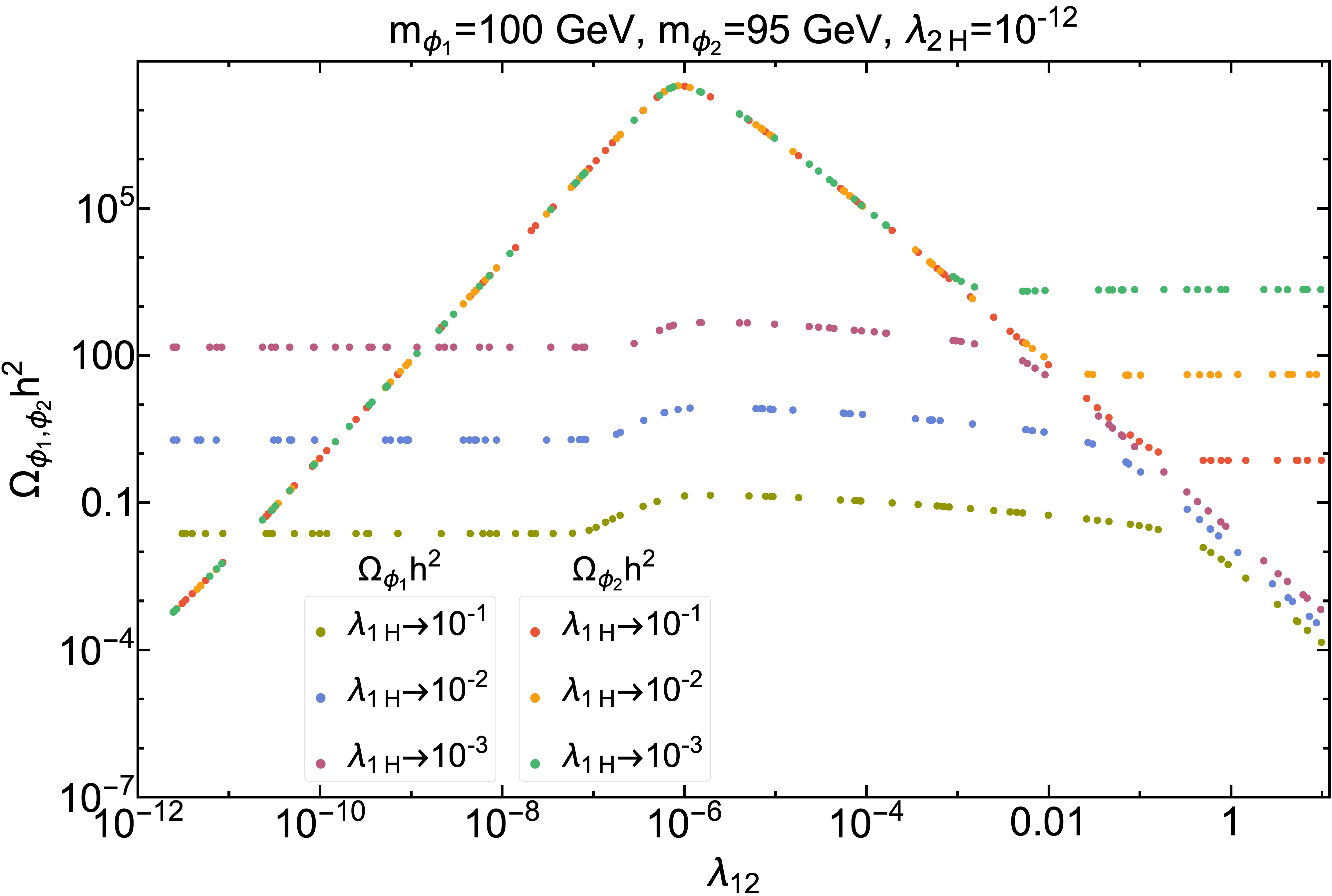}\label{fig:-model-m1gm2_relic_l1-l12}}~~
\subfloat[]{\includegraphics[width=0.45\linewidth]{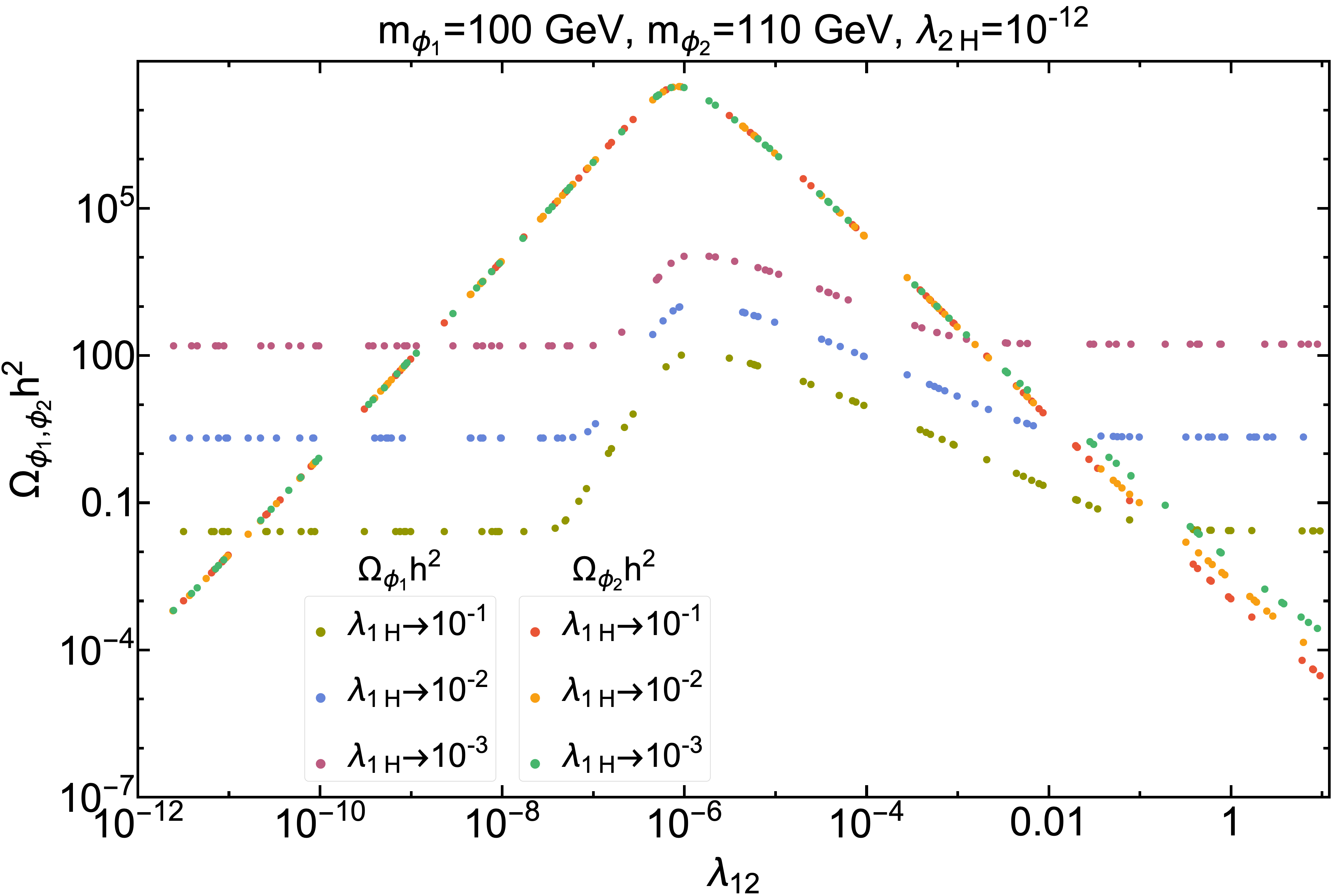}\label{fig:-model-m1lm2_relic_l1-l12}}
\caption{Variation of relic density with $\lambda_{12}$ for the the two component scalar DM model, where $\phi_1$ is {\tt WIMP} and $\phi_2$ is p{\tt FIMP} 
for $m_{\phi_1} > m_{\phi_2}$ (a) and $m_{\phi_1} < m_{\phi_2}$ (b). See figure insets and headings for other parameters kept fixed for the plot. }
\label{relic_model}
\end{figure*}

In figure \ref{relic_model}, we show the relic density of the DM components $\phi_1,\phi_2$ for a two component scalar DM scenario, as a function of the 
DM-DM interaction coupling $\lambda_{12}$. Both the hierarchies $m_{\phi_1} > m_{\phi_2}$ (a) and $m_{\phi_1} < m_{\phi_2}$ (b) are shown. We see 
a spectacularly similar feature to that of model independent analysis presented in fig. 1 of the main text, showing that the generic properties assigned to 
p{\tt FIMP} are generically valid.  

\section{Direct search prospect of p{\tt FIMP} via {\tt WIMP} loop}
\begin{figure}[htb!]\large
\centering
\begin{tikzpicture}[baseline={(current bounding box.center)},style={scale=0.9, transform shape}]
\begin{feynman}
\vertex (a);
\vertex[ right=1cm of a] (a2){\(\phi_2\)};
\vertex[ left=1cm of a] (a1){\(\phi_2\)}; 
\vertex[below=2cm of a] (c); 
\diagram* {(a1) -- [ scalar, arrow size=0.7pt, style=black] (a)-- [ scalar, arrow size=0.7pt, style=black] (a2),(a) -- [scalar, edge label={\(\rm \color{black}{h }\)},style=gray!50] (c) };
\end{feynman}
\end{tikzpicture}  +  
\begin{tikzpicture}[baseline={(current bounding box.center)},style={scale=0.9, transform shape}]
\begin{feynman}
\vertex (a);
\vertex[right=1cm of a] (a2){\(\rm \phi_2\)};
\vertex[left=1cm of a] (a1){\(\rm\phi_2 \)}; 
\vertex[below=1cm of a] (c); 
\vertex[below=1cm of c] (d); 	
\diagram* {(a1) -- [ scalar, arrow size=0.7pt] (a),(a) -- [ scalar, arrow size=0.7pt] (a2),(a) -- [scalar, half left, scalar, arrow size=0.7pt, style=red, edge label={\(\rm\color{black}{\phi_1 }\)}] (c) [crossed dot] -- [scalar, arrow size=0.7pt, half left, style=red, edge label={\(\rm\color{black}{\phi_1 }\)}] (a),(c) -- [scalar, style=gray!50, edge label={\(\rm\color{black}{h }\)}] (d)};
\end{feynman}
\end{tikzpicture}  +  
\begin{tikzpicture}[baseline={(current bounding box.center)},style={scale=0.9, transform shape}]
\begin{feynman}
\vertex (c);
\node[above=2cm of c, crossed dot, style=black](a);
\vertex[right= 1cm of a] (a2){\(\phi_2\)};
\vertex[left=1cm of a] (a1){\(\phi_2\)}; 
\diagram* {(a1) -- [ scalar, arrow size=0.7pt, style=black] (a) -- [ scalar, arrow size=0.7pt, style=black] (a2),(a) -- [scalar, edge label={\(\rm \color{black}{h }\)},style=gray!50] (c) };\end{feynman}
\end{tikzpicture}
\caption{The Feynman diagrams for p{\tt FIMP} ($\phi_2$) to interact with SM via Higgs. Sum of these tree (left), 1-loop (middle) and counter (right) 
diagram gives the total renormalized $\lambda_{h\phi_2\phi_2}$ coupling at 1-loop level.}
\label{dm-dd}
\end{figure}
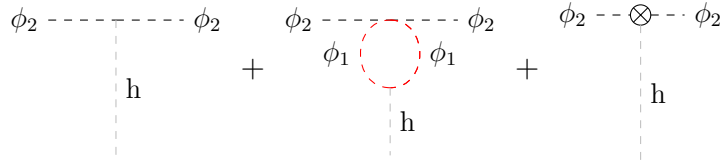
We will now provide the details of direct search prospect of the p{\tt FIMP} ($\phi_2$) in the two component scalar singlet model. 
The renormalized amplitude is given by (see \autoref{dm-dd}),
\begin{align}
\nonumber\Gamma_{h\phi_2\phi_2}&\,=\, \Gamma_{h\phi_2\phi_2}^{\rm tree}+\Gamma_{h\phi_2\phi_2}^{\rm 1-loop}+\Gamma_{h\phi_2\phi_2}^{\rm counter}\,,\\&
\,=\, -i \lambda_{h\phi_2\phi_2}+\Gamma_{h\phi_2\phi_2}^{\rm 1-loop}-i\delta_{\lambda_{h\phi_2\phi_2}}\,.
\end{align}
Here, 
\begin{eqnarray}
\Gamma_{h\phi_2\phi_2}^{\rm 1-loop}\left(q_h\right)\,=\, \frac{1}{2}\mu^{4-d}\left(-i\lambda_{\phi_1\phi_1\phi_2\phi_2}\right)\left(-i\lambda_{h\phi_1\phi_1}\right)\int \frac{d^dk}{(2\pi)^d}\frac{i}{k^2-m_{\phi_1}^2}\frac{i}{(k-q_h)^2-m_{\phi_1}^2}\,.\label{loop-1}
\end{eqnarray}
The transfer momentum $q_h$ is momentum transfer from initial to final state particles and $1/2$ is the symmetry factor due to the scalar loop. 
The dimension regularization parameter $\mu$ (mass scale) is introduced to keep couplings dimensionless in $d=4-2\epsilon$ dimension when $\epsilon\to0_+$. 
Now the 1-loop amplitude in \autoref{loop-1} becomes,
\begin{align}
\nonumber\Gamma_{h\phi_2\phi_2}^{\rm 1-loop}\left(q_h^2\right)\,=\, &\frac{1}{2}\left(-i\lambda_{\phi_1\phi_1\phi_2\phi_2}\right)\left(-i\lambda_{h\phi_1\phi_1}\right)\frac{-i}{16\pi^2}\int_0^1 dx\left(\frac{4\pi\mu^2}{\Delta}\right)^{\epsilon}\Gamma\left(\epsilon\right)\hspace{1cm}{\rm where},~\Delta=m_{\phi_1}^2-x(1-x)q_h^2\,,\\\nonumber\,
=\,& \frac{1}{2}\left(-i\lambda_{\phi_1\phi_1\phi_2\phi_2}\right)\left(-i\lambda_{h\phi_1\phi_1}\right)\frac{-i}{16\pi^2}\int_0^1 dx\left(\frac{1}{\epsilon}-\gamma_E+\ln\left[4\pi\mu^2\right]-\ln\left[\Delta\right]\right)\,,\\\,
=\,& \frac{1}{2}\left(-i\lambda_{\phi_1\phi_1\phi_2\phi_2}\right)\left(-i\lambda_{h\phi_1\phi_1}\right)\frac{-i}{16\pi^2}\int_0^1 dx\left(\frac{1}{\epsilon}-\gamma_E+\ln\left[4\pi\mu^2\right]-\ln\left[m_{\phi_1}^2-x(1-x)q_h^2\right]\right)\,.
\label{loop-2}
\end{align}
\begin{figure*}[hptb!]
\includegraphics[width=0.4\linewidth]{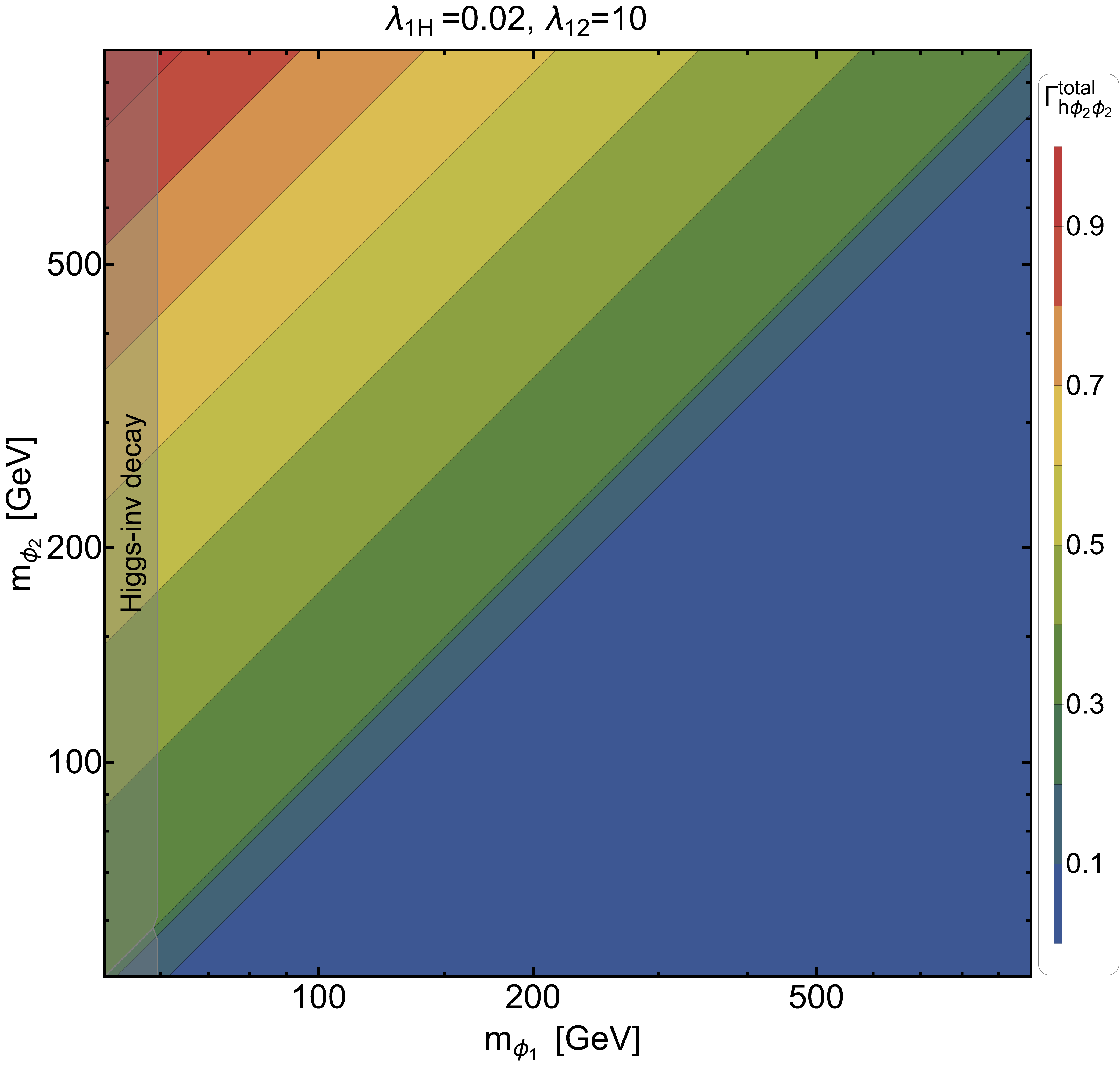}
\caption{Contour plot in $m_{\phi_1}-m_{\phi_2}$ plane to represent the absolute 1-loop amplitude 
$\rm|\Gamma_{h\phi_2\phi_2}^{{total}}|_{t\to 0}$ for some chosen values of $\lambda_{1H}$ and $\lambda_{12}$ as mentioned in the figure heading.}\label{fig:scalar_loop-amp}
\end{figure*}

There is an ambiguity of choosing the renormalisation scale, its unknown essentially. The direct search cross-section depends on the choice. For, $q_h^2\to 0$, 
the cross-section is vanishingly small \cite{DiazSaez:2021pfw}, however if we choose $q_h^2=4m_{\phi_2}^2$, a scale relevant for the p{\tt FIMP} production, the contribution is substantial \cite{Abe:2015rja}.
So we choose $q_h^2=4m_{\phi_2}^2$ as the renormalization condition setting point and calculate the counter term which removes the divergence from 1-loop amplitude 
$\Gamma_{h\phi_2\phi_2}^{\rm 1-loop}$ to yield,
\begin{align}
\Gamma_{h\phi_2\phi_2}^{\rm total}\left(q_h^2\right)\,=\,&  -i\lambda^{\rm relic}_{h\phi_2\phi_2}-\frac{i}{32\pi^2}\left(\lambda_{\phi_1\phi_1\phi_2\phi_2}\lambda_{h\phi_1\phi_1}\right)\int_0^1 dx\ln\left[\frac{m_{\phi_1}^2-x(1-x)4m_{\phi_2}^2}{m_{\phi_1}^2-x(1-x)q_h^2}\right]\,.
\label{loop-3}
\end{align}
Therefore, using \autoref{loop-3} the effective spin-independent DM-nucleon direct-detection cross-section in zero transfer momentum $(q_h^2=t\to 0)$ limit turns out to be,
\bea
\sigma_{{\phi_2}_{\rm eff}}^{\rm SI}\,=\,\frac{\Omega_{\phi_2}h^2}{\Omega_{\phi_1}h^2+\Omega_{\phi_2}h^2}\frac{\mu_n^2m_n^2}{4\pi v^2m_{\phi_2}^2}\frac{f_n^2}{m_h^4}|\Gamma_{h\phi_2\phi_2}^{\rm total}|^2_{t\to 0}\,.
\label{sigma-dd}
\eea
In the above, $\mu_n=\frac{m_nm_{\phi_2}}{m_n+m_{\phi_2}}$ where $m_n$ is the nucleon mass, $f_n=\frac{2}{9}+\frac{7}{9}\displaystyle\sum_{u,d,s}f_{T_q}^n$ with $f_{T_u}^{p(n)}=0.018(0.013),~f_{T_d}^{p(n)}=0.027(0.040),$ and $f_{T_s}^{p(n)}=0.037(0.037)$ \cite{Abe:2015rja}. We use \autoref{sigma-dd} for p{\tt FIMP} direct search cross-section while scanning the 
parameter space as shown in the main text.

\twocolumngrid

\end{document}